\documentclass[useAMS,usenatbib,usegraphicx]{mn2eadapt}
\usepackage[latin1]{inputenc}
\usepackage{amsmath,amsfonts,amssymb}
\usepackage{color}
\usepackage{times}

\newcommand{\Msun}{\ensuremath{\textrm{M}_{\odot}}}

\newcommand{\kms}{km\hspace{0.25em}s$^{-1}$}

\newcommand{\OI}{\mbox{O\hspace{0.25em}{\sc i}}}

\newcommand{\SiII}{\mbox{Si\hspace{0.25em}{\sc ii}}}

\newcommand{\CaII}{\mbox{Ca\hspace{0.25em}{\sc ii}}}

\newcommand{\FeII}{\mbox{Fe\hspace{0.25em}{\sc ii}}}
\newcommand{\FeIII}{\mbox{Fe\hspace{0.25em}{\sc iii}}}

\newcommand{\Nifs}{$^{56}$Ni}

\newcommand{\Mej}{M$_{\textrm{ej}}$}

\newcommand{\KE}{E$_{\rm K}$}

\newcommand{\vph}{$v_{ph}$}
\newcommand{\apj}{ApJ}

\newcommand{\mnras}{MNRAS}
\newcommand{\nat}{Nature}
\newcommand{\ssr}{Space Sci. Rev.}
\newcommand{\sci}{Sci.}

\newcommand{\eg}{e.g.\ }
\newcommand{\ie}{i.e.\ }

\def\gsim{\mathrel{\rlap{\lower 4pt \hbox{\hskip 1pt $\sim$}}\raise 1pt \hbox {$>$}}}
\def\lsim{\mathrel{\rlap{\lower 4pt \hbox{\hskip 1pt $\sim$}}\raise 1pt \hbox {$<$}}}
\def\gtaprx {\lower .1ex\hbox{\rlap{\raise .6ex\hbox{\hskip .3ex
	{\ifmmode{\scriptscriptstyle >}\else
		{$\scriptscriptstyle >$}\fi}}}
	\kern -.4ex{\ifmmode{\scriptscriptstyle \sim}\else
		{$\scriptscriptstyle\sim$}\fi}}}
\def\ltaprx {\lower .1ex\hbox{\rlap{\raise .6ex\hbox{\hskip .3ex
	{\ifmmode{\scriptscriptstyle <}\else
		{$\scriptscriptstyle <$}\fi}}}
	\kern -.4ex{\ifmmode{\scriptscriptstyle \sim}\else
		{$\scriptscriptstyle\sim$}\fi}}}

\begin{document}

\title[SN\,2013dx]
{Modelling of SN\,2013dx associated with the low-redshift GRB130702A points to diversity in GRB/SN properties}

\author[P. A.~Mazzali et al.]{P. A. Mazzali$^{1,2}$
\thanks{E-mail: P.Mazzali@ljmu.ac.uk}, 
E.~Pian$^{3,1}$, F.~Bufano$^4$, C. Ashall$^5$\\
\\
  $^1$Astrophysics Research Institute, Liverpool John Moores University, IC2, 
       Liverpool Science Park, 146 Brownlow Hill, Liverpool L3 5RF, UK\\
  $^2$Max-Planck-Institut f\"ur Astrophysik, Karl-Schwarzschild Str. 1, 
  	D-85748 Garching, Germany\\
  $^3$INAF, Astrophysics and Space Science Observatory, Via P. Gobetti 101, 
  40129 Bologna, Italy \\
  $^4$INAF, Astrophysical Observatory of Catania, Via Santa Sofia 78, 95123 
  Catania, Italy \\
  $^5$Institute for Astronomy, University of Hawai'i at Manoa, 
  2680 Woodlawn Dr., Hawai'i, HI 96822, USA
}

\date{Accepted ... Received ...; in original form ...}
\pubyear{2021}
\volume{}
\pagerange{}

\maketitle
  
\begin{abstract} 
The properties of the broad-lined type Ic supernova (SN) 2013dx, associated with
the long gamma-ray burst GRB\,130702A at a redshift $z = 0.145$, are derived via
spectral modelling. SN\,2013dx was similar in luminosity to other GRB/SNe, with
a derived value of the mass of \Nifs\ ejected in the explosion of $\approx
0.4$\,\Msun. However, its spectral properties suggest a smaller explosion
kinetic energy. Radiation transport models were used to derive a plausible mass
and density distribution of the SN ejecta in a one-dimensional approximation.
While the mass ejected in the explosion that is obtained from the modelling
(\Mej\,$\approx 9$\,\Msun) is similar to that of all other well-studied 
GRB/SNe, the kinetic energy is significantly smaller (\KE\,$\sim
10^{52}$\,erg).  This leads to a smaller \KE/\Mej\ ratio, $\approx
10^{51}$\,erg/\Msun, which is reflected in the narrower appearance of the
spectral lines. While the low \KE\ does not represent a problem for the scenario
in which magnetar energy aids powering the explosion and the nucleosynthesis, it
is nevertheless highly unusual. SNe\,Ic with similar \KE\ have never been seen
in coincidence with a GRB, and no well-observed GRB/SN has shown similarly low
\KE\ and \KE/\Mej.
\end{abstract}


\section{Introduction}
\label{sec:introduction}

Although long-duration gamma-ray bursts (GRBs) are relatively common events
($\sim 1$/day), only rarely, roughly once per year or less, do they happen close
enough to the Earth that we can study the stellar death that accompanies them in
any detail \citep{gal98,hjorth03,levan16}.

The understanding of the significance of the broad-lined spectra that are so
characteristic of the Type Ic SNe that typically accompany GRBs\footnote{Type Ic
SNe are core-collapse events of stars stripped of their outer H and He layers.
See \citep{filipp97} for a spectral classification of SNe.} led to the
realisation that GRB/SNe are extremely energetic explosions \citep{iwa98,min00},
where material from the carbon-oxygen core of the progenitor star is ejected
with kinetic energies (\KE) of several $10^{52}$\,erg, \ie\ more than an order
of magnitude larger  than the so-called ``typical'' SN energy. In fact, the
prototypical GRB/SN\,1998bw had a \KE\ as high as $4 \times 10^{52}$\,erg in
spherical symmetry \citep{iwa98}, and other events that have been carefully
analysed have yielded similar values, from 
SN\,2003dh \citep[$3 \times 10^{52}$\,erg,][]{maz03} to 
SN\,2003lw \citep[$5 \times 10^{52}$\,erg,][]{maz06a}, and 
SN\,2016jca \citep[$4 \times 10^{52}$\,erg,][]{ashall19}.

The extremely high \KE\ is invariably accompanied by a rather large ejected mass
\Mej\,$ \sim 10$\,\Msun, and a high luminosity, which is the result of the
synthesis and radioactive decay of several tenths of a \,\Msun\ of \Nifs, again
an unusually high value for core-collapse SNe (GRB/SNe produce almost as much
\Nifs\ as SNe\,Ia). As GRB/SNe are all of type Ic \citep[with the possible
exception of the Type Ib SN\,2008D, which was accompanied by an X-ray
flash,][]{soderberg08,maz08a,modjaz09,tanaka09}, the ejected mass typically
reflects the mass of the carbon-oxygen core of the progenitor star, which scales
roughly with the mass of the star at birth. Such relations
\citep[\eg][]{nomhash88} suggest progenitor zero-age main sequence masses of
$\sim 35-50$\,\Msun. It is therefore natural to hypothesize that the large mass
of these core-collapse events is in some way responsible for the large \KE, or,
at least, that it is one of the driving factors. How this may exactly happen is
not  clear, however, as it is very unlikely that neutrinos can deposit so much
of their energy to the stellar envelope \citep{janka2016}. One current
hypothesis is that energy is extracted from a highly rotating, magnetised
proto-neutron star (a magnetar) that forms upon the collapse of the core
\citep{bucciantini2009,metzger2015}. Given the high mass of the progenitors, it
is more likely that the final result of the collapse is a black hole, but a
temporary neutron star phase may not be excluded, especially if the compact
object is born with a high rotation rate. A magnetar can store up to $\sim 2.5
\times 10^{52}$\,erg of rotational energy, which is sufficient to power every
GRB/SN if we assumed that the isotropic \KE\ derived from modelling on-axis
events is overestimated by a factor of 2-3 as a consequence of the asphericity
of the ejecta, which would be a natural consequence of the magnetic field
configuration \citep{Umcf2007}, and our privileged viewing angle
\citep{mae02,maz05}. How the energy is extracted is again unclear. What is clear
is that the energy must be extracted very quickly, in a matter of a second or
less, in order to power the explosion and lead to the synthesis of large amounts
of \Nifs\ \citep{suwa15,chen17}.  Although it is sometimes suggested that the
long-term decay of the magnetar can power the SN light curve
\citep[\eg][]{woosley10,kasbild10} this is not necessary for a GRB/SN, since we
know from nebular data \citep{maz01b,maeda07,maz07} that the radioactive decay
of \Nifs\ can power both the light curve and the nebular epoch emission, which
is dominated by lines of oxygen and iron. In the case of GRB/SNe magnetic energy
could energise the explosion, leading to the observed near-constant explosion
kinetic energy \citep{maz14}, and possibly contribute to the synthesis of \Nifs.

As the field is awaiting a theoretical breakthrough, the study and analysis of
individual events is certainly a worthwhile effort, both to confirm the trend
and to identify exceptions, should any be found, which may light the path to 
answering to the questions presented above. 

The case of SN\,2013dx is an intriguing one. The SN was associated with
GRB\,130702A, a rather normal long GRB \citep{singer13,delia15,toy16}. The SN
was as luminous as other GRB/SNe, and its spectrum showed broad lines
\citep{delia15}, but at a closer look line blending was not quite as extreme as
in textbook cases such as SN\,1998bw. We know that line blending is a proxy for
\KE\ \citep{maz17}. \citet{prenticemazz18} introduced a simple method to
estimate the relative \KE\ of SNe\,Ic using a count of the number of spectral
features in the optical spectral range near maximum: GRB/SNe show typically 3
features and are classified as Ic-3. SN\,2013dx, on the other hand, showed 4
features, with the Fe-dominated absorption at 4000-5000\,\AA\ split into two
components, and quite possibly also the \CaII\,IR triplet not blending with the
(weak) \OI\,7774 line, which suggests that the amount of material moving with $v
> 0.1 c$ was significantly smaller than in other GRB/SNe. Therefore, while most
previously studied GRB/SNe required only a small change to the original model
that was used to match SN\,1998bw, SN\,2013dx offers the opportunity to explore
a different range of parameters and to apply the experience we have gained from
modelling both GRB/SNe and other broad-lined SNe\,Ic that did not show an
accompanying GRB \citep[\eg][]{maz13,maz17}. 

In this paper, we analyse the spectra first presented by \citet{delia15} and
present a new spectrum taken with the ESO VLT about 6 months after explosion,
when the SN was expected to be in the nebular phase.  The data have been newly
decomposed, as discussed in Section \ref{sec:Data}. Spectral models were
computed using our well-established Montecarlo radiation transport code for SNe.
As discussed in Section \ref{sec:Method}, different explosion models were used
in order to determine as accurately as possible the values of \Mej\ and \KE.
Lacking reliable nebular data, we were forced to constrain the \Nifs\ mass using
only a combination of the abundances derived from the early-time modelling and a
light curve simulation. This was done using our Montecarlo light curve code, and
is discussed in Section \ref{sec:LC}. Having obtained a reasonable result from
modelling, we discuss its possible significance in Section \ref{sec:Disc}, and
conclude in Section \ref{sec:Conclusion}.

\section{Revised decomposition of spectra and light curves}
\label{sec:Data}

In this Section we present an updated decomposition of the optical counterpart
of GRB130702A into host galaxy, afterglow, and supernova contributions. This
improves on previous attempts by taking into account a larger dataset and an
accurate method for afterglow modelling \citep{ashall19}.  Although all
available optical photometry is used \citep{singer13,delia15,toy16}, our SN
spectral analysis only relies on the VLT+FORS spectra \citep{delia15}, which
have the best signal-to-noise ratio.

\subsection{Late-epoch spectrum and host galaxy}

\begin{figure} 
\includegraphics[width=90mm]{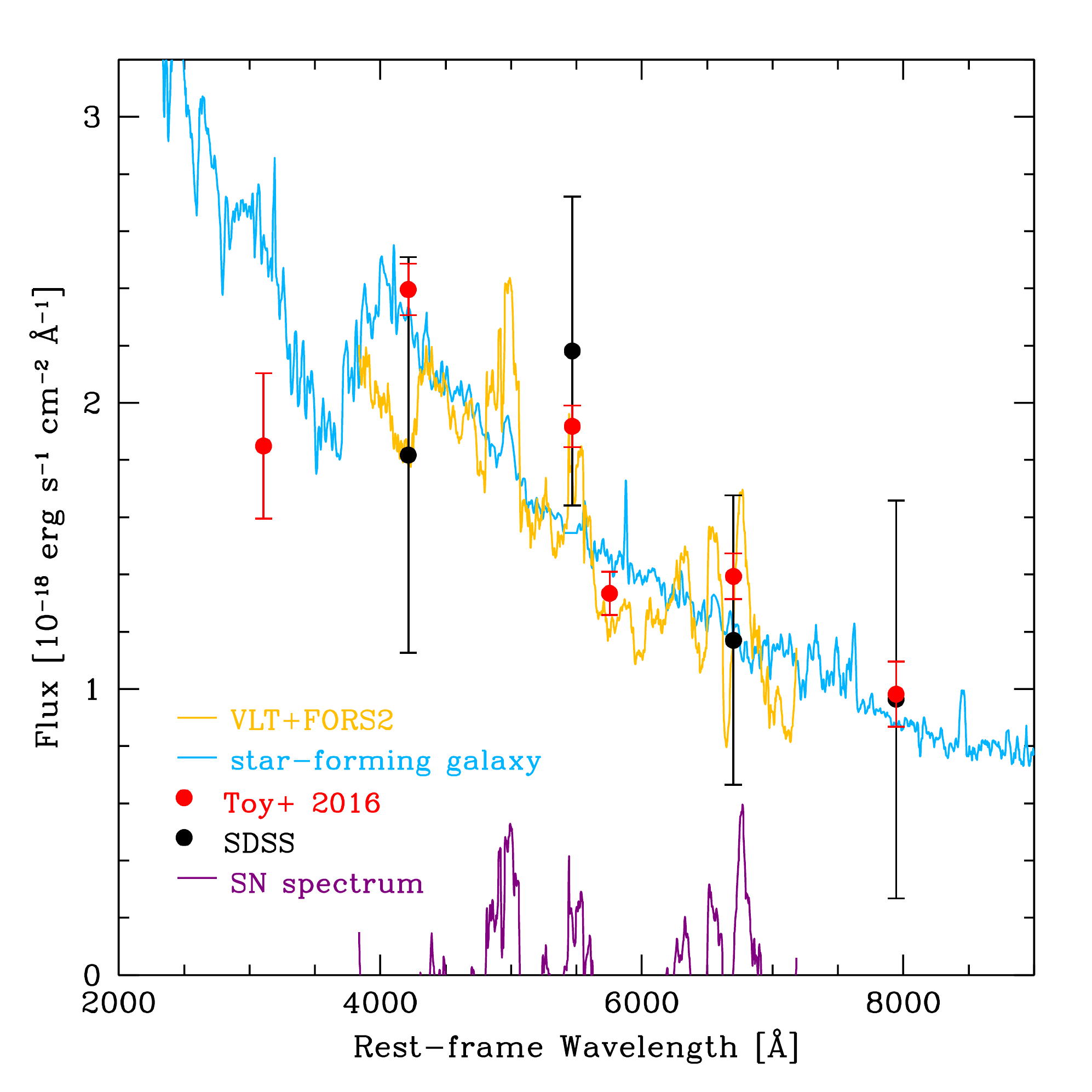}
\caption{VLT+FORS2 spectrum taken in Jan-Mar 2014 (yellow), cleaned of spurious
features, corrected for redshift ($z =0.145$), Galactic absorption (E(B-V)$ =
0.037$\,mag), smoothed with a boxcar of 50\,\AA, and scaled to match the
late-epoch flux measurements made with Keck in the {\it ugriz} filters
\citep[red circles, from][]{toy16}.  The SDSS {\it ugriz} photometry of the
host galaxy is also shown (black circles).  The template of a star-forming
galaxy with low intrinsic extinction (E(B-V) $< 0.1$\,mag), from
\citet{kinney96}, is shown for comparison (light blue).  The violet curve is
the residual SN spectral signal obtained after subtracting the star-forming
template spectrum from the VLT spectrum.  No identifiable nebular emission lines
are seen, notably [O I] 6300,6363.}
\label{fig:lateepochvltspec} \end{figure}

Late-epoch spectroscopic exposures with the VLT and FORS2 equipped with grism 
300V+GG435 were taken on 2014 Jan 22 (2x1800s), Mar 11 (2x1800s) and Mar 13
(4x1800s), i.e. 178, 220 and 222 rest-frame days after explosion, respectively,
in order to search for emission lines, the strongest spectral signature
expected from the SN in the nebular phase. The spectra were reduced with
standard methods.  No nebular emission lines appear to be superposed on to the
continuum, which is dominated by the flux of the host galaxy, as shown in Figure
\ref{fig:lateepochvltspec}.  In this figure we show the co-added VLT spectrum
scaled in flux to match the late-epoch photometry taken at 330 and 632 days
after explosion \citep{toy16}, and compare it to SDSS photometry and to the
template of a little-extinguished star-forming galaxy (Kinney et al. 1996; see
also Kelly et al. 2013).  

The  spectra are generally consistent with the photometry, with the exception of
the SDSS points at $\sim$\,4000 and 5500\,\AA, which have however large
errors.   The VLT spectrum thus represents exclusively the host galaxy. 
However, owing to its poor quality and limited wavelength extension, we
preferred to use the low-extinction star-forming template for subtraction of 
the host galaxy contribution from all spectra in our study.  This choice is
justified by the fact that long GRBs are known to be generally hosted by
galaxies with high star-formation rate and in this specific GRB there is no
evidence of significant intrinsic extinction.  For the NIR photometry, we
extrapolated the template spectrum to longer wavelengths using a power-law.

%{\bf In order to estimate an upper limit on nebular line emission, we have
%subtracted  the star-forming template also from this late-epoch VLT spectrum  
%(see violet curve in Figure \ref{fig:lateepochvltspec}).  From the subtracted
%spectrum we evaluated an upper limit of $6.5 \times 10^{-18}$ erg s$^{-1}$
%cm$^{-2}$ on the intensity of the [O I]6300,6363 emission, typically the
%strongest core-collapse SN  transition in nebular phase, corresponding to a
%luminosity lower than $3.4 \times 10^{38}$ erg s$^{-1}$.}

\subsection{Multi-wavelength afterglow}

In order to isolate the SN component, previous authors estimated the afterglow
contribution assuming it is produced by synchrotron radiation in a relativistic
shock that interacts with a uniform external medium, as prescribed by the
classical fireball model \citep[$f(t) \propto t^{-\alpha} \nu^{-\beta}$,][and
references therein]{zhangmesz04,racusin09}.  The energy distribution of the
relativistic particles follows a power-law of the form $dN/dE \propto E^{-p}$. 
The characteristic  frequency of the particles that lose a substantial fraction
of their energy to synchrotron radiation (cooling frequency) is indicated with 
$\nu_c$.

While \citet{delia15} limited their analysis to the optical light curve,
\citet{singer13} and \citet{toy16} considered also the {\it Neil Gehrels Swift
Observatory} ({\it Swift}) XRT light curve (0.3-10 keV) and tried a simultaneous
fit of both X-ray and optical data within the above model.  However, they did
not take into account the first $r$-band photometric point, and the afterglow
decay prior to the time break that they measure to occur at 1.17 days.

Here we propose a model of the optical and X-ray afterglow of GRB130702A based
on  classical fireball theory that consistently takes into account the observed
spectral slopes, decay indices and $r$-band light curve steepening.  In doing
so, we adopted the same approach used in the case of the multi-wavelength
afterglow of GRB161219B \citep{ashall19}, which required the injection of a
refreshed blast wave starting within a few hours of the main GRB event, and used
that same formalism to evaluate the model parameters. 

We downloaded from the {\it Swift} archive the most up-to-date XRT light curve
of GRB130702A, which extends to more than 100 days after explosion
(https://www.swift.ac.uk/xrt\_curves/00032876/).  This is consistent with a
single power-law of index $\alpha = 1.21 \pm 0.02$.  The X-ray spectrum,
de-absorbed for a Galactic $N_{HI} = 1.83 \times 10^{20}$ cm$^{-2}$ and a poorly
constrained intrinsic $N_{HI} \simeq 6 \times 10^{20}$ cm$^{-2}$, is well fitted
by a single power-law with photon index $\Gamma = \beta_X + 1 = 1.82 \pm 0.12$
(http://www.swift.ac.uk/xrt\_spectra/00032876/). The optical spectrum taken
$\sim 1$ day after explosion was corrected for Galactic \citep[$A_V$ = 0.116
mag, ][]{sf11} as well as intrinsic absorption \citep[$A_V$ = 0.10
mag,][]{toy16}. It can then be represented by a single power-law with
$\beta_{opt} = 0.52 \pm 0.19$ \citep{toy16}. This is flatter than the X-ray
spectral slope, although still marginally consistent with it.  The optical light
curves (except in the $r$-band) were monitored starting $\sim 1$ day after
explosion. They decay as single power-laws with index $\alpha \simeq 1.1-1.2$
between day 1 and day 4 \citep{singer13,toy16}.  The $r$-band light curve, which
was monitored starting $\sim 4$ hours after explosion, cannot be described by a
single power-law. It requires a broken power-law with an early decay of $\alpha
= 0.57$ and a steepening after a few days \citep{singer13,toy16}, which appears
to be larger ($\Delta\alpha \approx 0.7$) than a cooling break ($\Delta\alpha =
0.25$).  At epochs later than 4 days all optical light curves exhibit an extra
component, which has been interpreted as the emergence of supernova light
\citep{singer13,delia15,toy16,volnova17}.

If the $r$-band light curve steepening observed a few days after explosion is
due to jet geometry and the consequent coming into view of the side of the jet,
it must occur achromatically.  According to fireball theory, if $p = 2.64$ as
derived from $\beta_X$ assuming that $\nu_X < \nu_c$, the X-ray light curve
(which is unaffected by the SN light) should decay as $t^{-2.64}$ after the
break, which is inconsistent with the {\it Swift}/XRT data. A fit to the X-ray
light curve with a broken power-law cannot reproduce a post-break $t^{-2.64}$
decay for any value of the break time. Therefore, we assume hereafter $\nu_X >
\nu_c$, which appears also more realistic (it is unlikely that $\nu_c$ is larger
than the X-ray frequency at a few days after explosion). 

If $\nu_X > \nu_c$, $p = 2 \beta_X = 1.64$, i.e. $p < 2$, which in principle
requires that we adopt the parameter dependencies that are appropriate for a
"flat" electron-energy law \citep{dai01,zhangmesz04}.  However, the $p < 2$
fireball prescription for frequencies larger than $\nu_c$ does not yield a good
fit to the X-ray light curve: the time indices before and after the break are
too steep for any break between 1 and a few days.  We then resorted to the
hypothesis and formalism developed in \citet{ashall19}: although formally the
electron-energy law is highly energetic ($p < 2$), which causes rapid cooling
and therefore steep decay indices, the external shock that produced the
afterglow is not an impulsive blast, but is refreshed a few hours after the
explosion, so that the added energy slows down the decay rates.  Under this
assumption, we used $p = 1.64$, but applied to the X-ray and optical afterglow
the closure relations that are appropriate for the case $p > 2$
\citep{zhangmesz04}.  

\begin{figure*} 
\includegraphics[width=180mm]{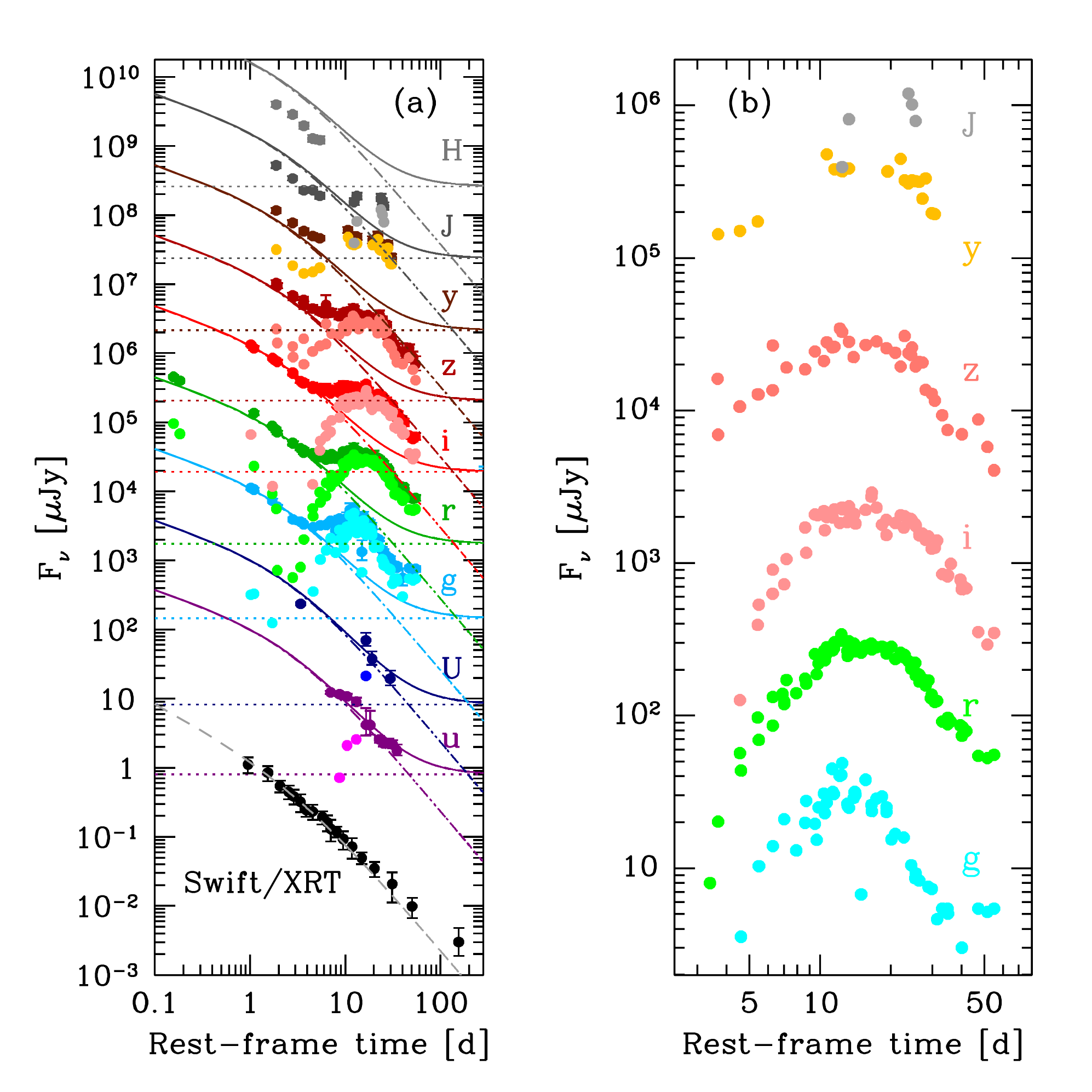}
\caption{a) Multi-wavelength light curves of the GRB130702A counterpart in
rest-frame (filled circles), constructed based on our data as well as data from
the literature \citep{singer13,delia15,toy16}.  With the exception of the
u band, which is in actual flux units, all optical and NIR light curves are
shifted vertically by increasing powers of 10.  The data were k-corrected with
our spectra (optical filters) and de-absorbed for Galactic extinction with
E(B-V) = 0.037 mag, using the extinction curve of \citet{cardelli89}.  The
dotted horizontal lines represent the host-galaxy contribution in the various
bands; the dot-dashed curves represent the optical afterglow modelled with a
steepening power-law (see text). The solid curve is the sum of the afterglow and
host-galaxy components. The subtraction of these two components from the
observed points corresponds to the SN light curves (represented with circles of
lighter hue than the observed points in the corresponding band; in H-band the
subtraction does not return any positive SN residual signal). The errors on the
SN points range from 10 to 15 per cent around maximum to 50 per cent at early
epochs, when the contribution of the afterglow is larger. They are not reported
for clarity. The black filled circles are the Swift/XRT light curve at 1 keV;
the best-fitting afterglow model at this energy is reported as a grey dashed
curve. 
b)~Same as (a), with SN light curves only, and limited to the bands where the 
subtraction returns a significant SN signal. With the exception of the g band, 
which is in actual flux units, all light curves are shifted vertically by 
increasing powers of 10.}
\label{fig:mwllccurves}
\end{figure*}

The multi-wavelength light curves, corrected for Galactic and intrinsic
extinction and k-corrected using the XRT and VLT spectra, are shown in Fig.
\ref{fig:mwllccurves}, left-hand panel, before and after host galaxy and
afterglow subtraction, and with model curves.  In Fig.
\ref{fig:mwllccurves}, right-hand panel, we reported only the subtracted SN
curves in the optical and NIR bands where the SN signal is significant.  The
simultaneous best fit of the X-ray and all optical light curves, carried out via
a $\chi^2$-minimization routine, is obtained for a time break of $\approx 3$
days.  The model X-ray decay index before the time break, $\alpha_X = 0.73$, is
apparently too steep for the optical $r$-band light curve, which suggests that
$\nu_c$ is located between the optical and X-ray domain (the best-fit value at 1
day is $\nu_c = 1.5 \times 10^{15}$ Hz). This is also consistent with the slope
of the optical spectrum  ($\beta_{opt} \sim 0.5$) determined by \citet{toy16}.  

If the extra luminosity input required by our approach is modelled as $L(t)
\propto t^{-q}$, the $q$ parameter can be derived as in \citet[][see their
Appendix]{ashall19} to be $q \simeq 0.84$, which is very similar to that
determined by \citet{ashall19} for GRB161219B, and conveniently smaller than 1,
so that the integrated luminosity increases with time and efficiently
re-energizes the shock until a cutoff is reached.

\subsection{Supernova component}
\label{sncomponent}

\begin{figure} 
\includegraphics[width=90mm]{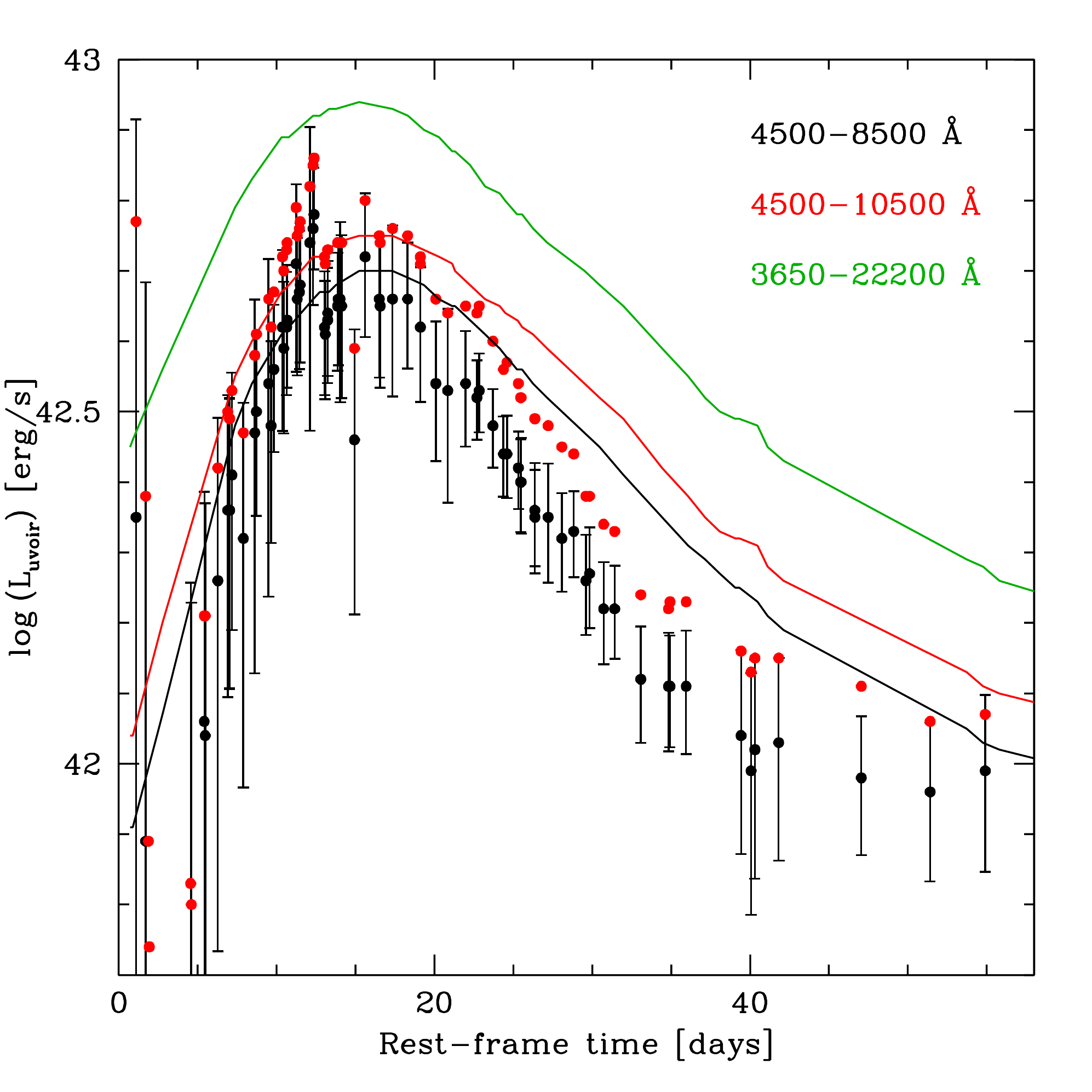}
\caption{Pseudo-bolometric light curves of SN2013dx integrated over the range 
4500-8500\,\AA\ (black circles) and 4500-10500\,\AA\ (red circles); the
uncertainties in the latter points are omitted for clarity.  The initial points
carry a large uncertainty and do not support any significant claim for the
presence of an extra component (e.g. shock-breakout). For comparison, the light
curves of SN\,1998bw computed in the same ranges are shown as solid curves with
the same respective colours, as is the light curve of SN\,1998bw computed over
the range 3650-22200\,\AA\ range (green curve).  The data of SN\,1998bw were
corrected for Galactic extinction E(B-V) = 0.052 mags; a distance of 35.1 Mpc
was assumed.}
\label{fig:bollcs_13dx_98bw}
\end{figure}

By subtracting the host galaxy and the afterglow contributions from the observed
optical light curves we obtained the SN light curves (shown in lighter tones in
Fig. \ref{fig:mwllccurves}).  We used the monochromatic light curves in which
the SN was best detected and monitored to construct a rest-frame
pseudo-bolometric light curve in the range 4500-10500\,\AA.  In order to do this
we interpolated the {\it g,r,i,z,y} fluxes at intervals of 1 day, integrated the
resulting spectral flux distribution over the corresponding wavelength range
(4825-10200\,\AA), added flux shortwards and longwards of the wavelength
boundaries of this range to 4500 and 10500\,\AA, respectively, by assuming a
flat flux power-law with respect to wavelength, and remapped it to the exact
epochs of observation.  We then computed the pseudo-bolometric luminosity using
$z = 0.145$ and assuming $H_0 = 73$\,\kms\,Mpc$^{-1}$ \citep{riess16} and a
concordant cosmology. This is shown in Fig. \ref{fig:bollcs_13dx_98bw}. 

The curve shape and luminosity are consistent with those published by 
\citet{delia15} and \citet{toy16}, considering the different assumptions on the afterglow
behaviour and subtraction methods, and the slightly different wavelength ranges
and cosmologies that were adopted.

We compared the light curve of SN\,2013dx with that of SN\,1998bw, which is
better sampled in the NUV and NIR \citep{gal98,patat01}.  First we compared the
light curves of both SNe in the range 4500-8500\,\AA\  (constructed as described
above), where they are both well observed: these light curves match well near
maximum (Fig. \ref{fig:bollcs_13dx_98bw}), although they deviate thereafter,
presumably owing to the larger ejecta mass of SN\,1998bw, but also possibly
because of the presence of a high density inner core in the ejecta of SN\,1998bw
\citep{mae03}.  Then we compared the light curves of the two SNe computed over
the same wavelength range, 4500-10500\,\AA, which is the range over which
SN\,2013dx has the best coverage.  The match is still very good, which
underlines the similarity of the two SNe and justifies using SN\,1998bw as a
comparison template for SN\,2013dx.  The light curve of SN\,1998bw in the full
range of coverage, 3650-22200\,\AA, differs by 0.18\,dex from that in the range
4500-10500\,\AA. 

\begin{figure} 
\includegraphics[width=90mm]{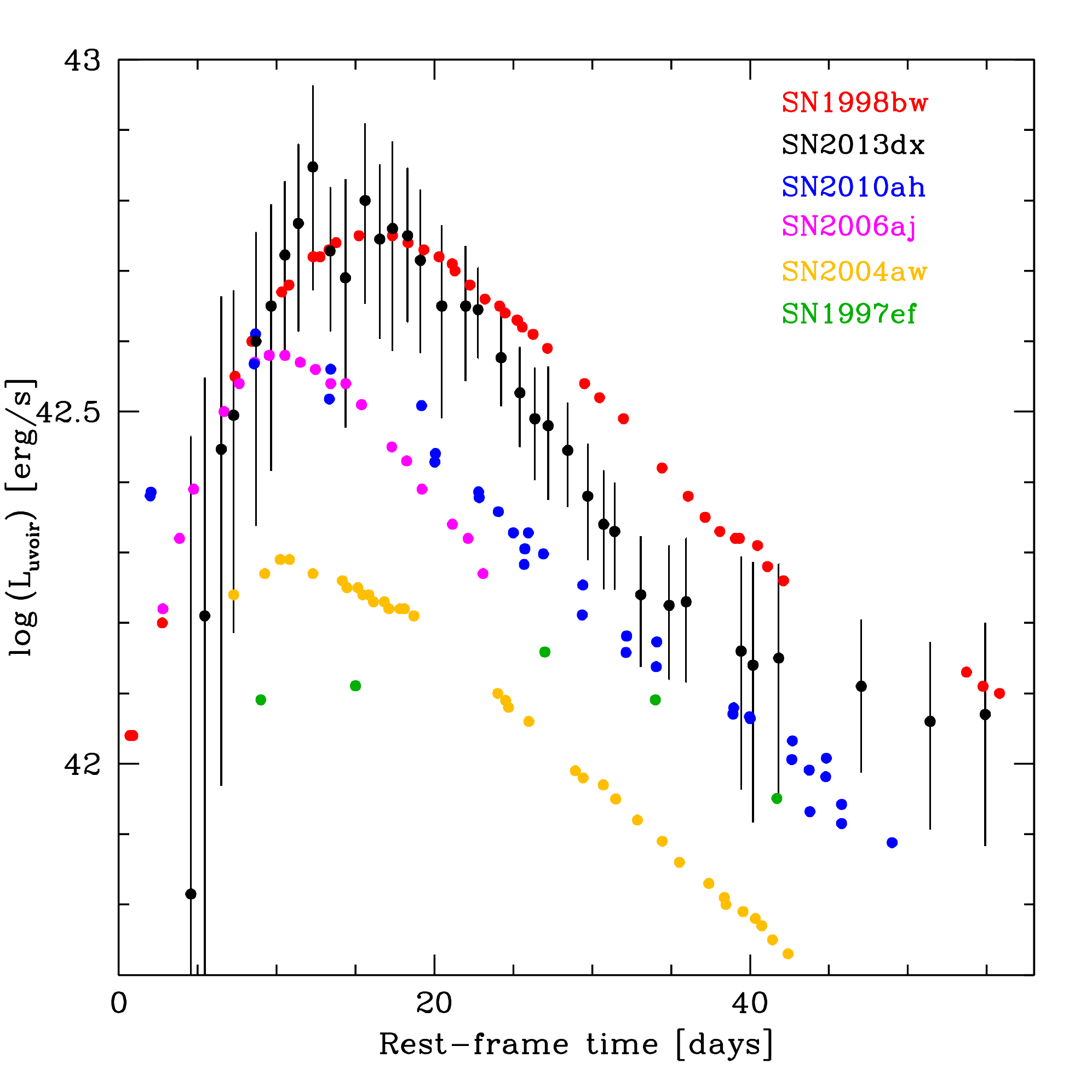}
\caption{Pseudo-bolometric light curves of well-observed stripped-envelope
core-collapse SNe, including GRB-SNe\,1998bw \citep{gal98}, and SN2013dx, XRF-SN
2006aj \citep{pian06,ferrero06}  and  SNe\,1997ef, 2004aw and 2010ah, all
broad-lined SNe that were not accompanied by  GRBs \citep{iwa00,maz13,maz17}.
The light curves were computed in the same wavelength range (4500-10500 \AA),
with the exception of SN\,1997ef which was only observed in the V-band. For
clarity, errors are only reported for SN2013dx.  Data of  SN\,2013dx earlier
than 4 days are omitted as they are affected by a large uncertainty (see Fig. 
\ref{fig:bollcs_13dx_98bw}), while the data that are shown have been binned so
that only one data point per day is shown.}
\label{fig:LCcomp}
\end{figure}

A comparison of SN\,2013dx with various broad-lined Ic SNe both accompanied
and not accompanied by high-energy transients shows that SNe\,Ic with different
properties can exhibit broad light curves (Fig. \ref{fig:LCcomp}). However,
GRB/SNe (2013dx and 1998bw in this figure) have larger luminosities.

%A comparison of both SN\,1998bw and SN\,2013dx with the broad-lined Ic SN\,2010ah \citep{maz13} that was not accompanied by a high-energy transient shows that  SNe\,Ic with different properties can exhibit broad light curves (Fig. \ref{fig:LCcomp}). However, the two GRB/SNe shown here have broader light curves than the non-GRB SN\,2010ah.  

Finally, in order to isolate the SN spectra, we subtracted the modelled
synchrotron optical afterglow and the host galaxy template spectrum from the
observed VLT spectra, corrected for Galactic extinction. The final decomposed
spectra are presented in the next Section.

\section{Spectral Modelling}
\label{sec:Method}

The results presented in \citet{delia15} were based on a simple appoximation
\citep{arnett82}. In particular, they were reliant on approximate measurements
of the ejecta velocity as estimated from the position of a very weak line that
was identified as \SiII\,6347, 6371. However, because of line blending, and the
quality of the spectra \citep[see Fig. 3 in][]{delia15}, this velocity is very
difficult to measure, and the results based on it are consequently highly
uncertain. Additionally, a line such as \SiII\,6347, 6371, which is
intrinsically strong, may form well above the pseudo-photosphere, as is well
known \citep[see, \eg][]{m00}. Given the unexpected spectral properties of
SN\,2013dx, it is interesting to make an attempt to determine the main
parameters of the SN more accurately than previously done, using a more complex
and physical approach. Therefore, here we use a full spectral modelling, which
allows us to estimate the photospheric velocity \vph\ rather than just the line
velocity. 

We followed the procedure we often adopted before when modelling SN data. We
used our Montecarlo radiation transport code for SNe, which is based on the
principles outlined in \citet{ml93,l99,m00}. The code uses the
Schuster-Schwarzschild approximation, assuming that the SN luminosity is emitted
with a black-body spectrum from the surface of a pseudo-photosphere whose
position moves inwards in mass coordinates with time as the ejecta expand and
thin out. Energy packets - which are representative of photons - are injected
into the SN envelope, and their propagation is tracked using a Montecarlo
scheme. Packets can interact with the gas in the SN envelope via line absorption
or electron scattering. In the case of line absorption, the excited electron is
allowed to de-excite through a number of randomly selected allowed downward
paths, such that new photons of different wavelength can be emitted. This 
automatically takes into account the process of line branching, which is
essential for the formation of SN spectra \citep{m00}. Electron scattering, on
the other hand, simply causes a packet to change its direction of travel.
However, by doing so, it increases the residence time of energy packets in the
SN ejecta, thereby increasing also the probability of line absorption. Radiative
equilibrium is enforced, as none of the interactions that are considered results
in the loss or gain of energy by the gas or the radiation field when a
sufficiently large number of interactions are taken into account. Deviation from
local thermodynamic equilibrium (LTE) in the low density, radiation-filled
environment of the SN envelope is taken into account by adopting the nebular
approximation \citep{abblucy85,ml93} for the level population and ionization.
This approximation is known to be in excellent agreement with detailed results
of a fully  non-LTE (nLTE) code \citep{pauld96}. After the last scattering
event, packets that escape are binned according to their frequency in a
Montecarlo spectrum, which is characterised by noise because of the method
adopted. A more accurate spectrum can be obtained if a post-iteration is
performed when emissivities are computed based on the final occupation numbers,
and the formal integral solution of the transfer equation is performed. These
are the spectra that are shown here. 

The Montecarlo code requires as input the emerging SN luminosity $L$, a
photospheric velocity \vph, and a time from explosion $t$, from which a
photospheric radius is computed. Density in a SN decreases with radius,
depending on the properties of the progenitor star and the energy of the
explosion. Thus a density distribution with radius, a so-called explosion model,
has to be used, which allows the homologously expanding ejecta to be re-mapped
in density to the time that is required for the spectral calculation. Abundances
are also a function of radius. In our code they can be modified as required to
optimise the fit to the spectrum. Our code has been applied to different types
of SNe. In \citet{maz08b} we discussed how errors on different quantities can be
estimated. This requires looking at the impact that changing each quantity has
on the highly non-linear problem that we are solving. In \citet{ashall19} we
discussed how a broad-lined SN\,Ic can be analysed and how information can be
extracted from the data and the fits, including indirect information about the
three-dimensional distribution of the ejecta and the abundances within them, if
sufficiently early spectra are available. 

The ejected mass \Mej\ and the explosion kinetic energy \KE\ of well-studied
GRB-SNe have so far clustered around values of $\sim 10$\,\Msun\ and $3 \times
10^{52}$\,erg, respectively. The energy refers to the isotropic value, which is
probably somewhat overestimated, because in GRB/SNe we are likely to observe the
fastest-moving ejecta in an aspherical explosion \citep[\eg][]{ashall19}.
However, other than the possible requirement of a large progenitor mass, and the
hypothesis that the limiting magnetar energy sets an upper limit to the energy,
these values are not motivated by any stringent physical argument, unlike for
example the \Mej\ and \KE\ of SNe\,Ia. Therefore, selecting a reasonable model
is the most important task in a modelling effort, and the most laborious.  

The width of the spectral features is normally used as a guideline for the
estimate of \KE, while the shape of the light curve can be used to estimate
\Mej. In this sense, as we already noticed in \citet{delia15}, SN\,2013dx
resembles spectroscopically  energetic SNe\,Ic that were not accompanied by an
observed GRB, such as SNe 2010ah \citep{maz13} or 1997ef \citep{min00}, rather
than any of the well-observed GRB/SNe. We showed in \citet{maz17} that the slope
of the outer density profile directly affects the appearance of the spectra, in
particular the line blending at the earliest times, when line formation occurs
in those outermost layers, which by virtue of their high velocity are capable of
carrying a significant fraction of the total \KE. We applied the same
methodology here. At the same time, the light curve of SN\,2013dx is broader
than that of - say - SN\,2010ah (see Fig. \ref{fig:LCcomp}), which indicates a
large \Mej, possibly of the order of 8-10\,\Msun. 

\begin{figure} 
\includegraphics[width=90mm]{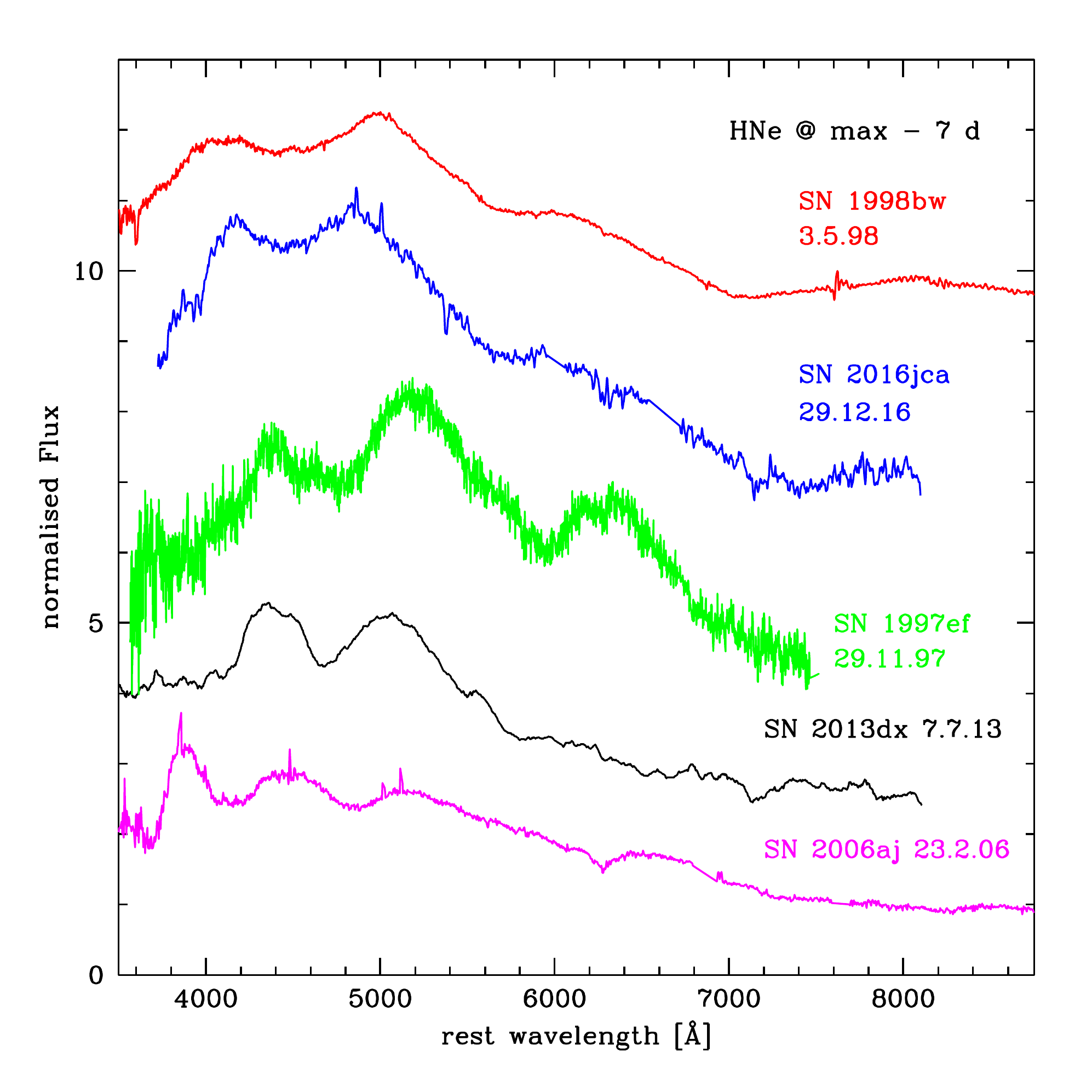}
\caption{The spectrum of SN\,2013dx taken $\approx 1$ week before maximum light 
compared to those of other broad-lined SNe\,Ic, both with and without a GRB, 
obtained at a similar epoch.}
\label{fig:speccomp_premax}
\end{figure}

\begin{figure} 
\includegraphics[width=90mm]{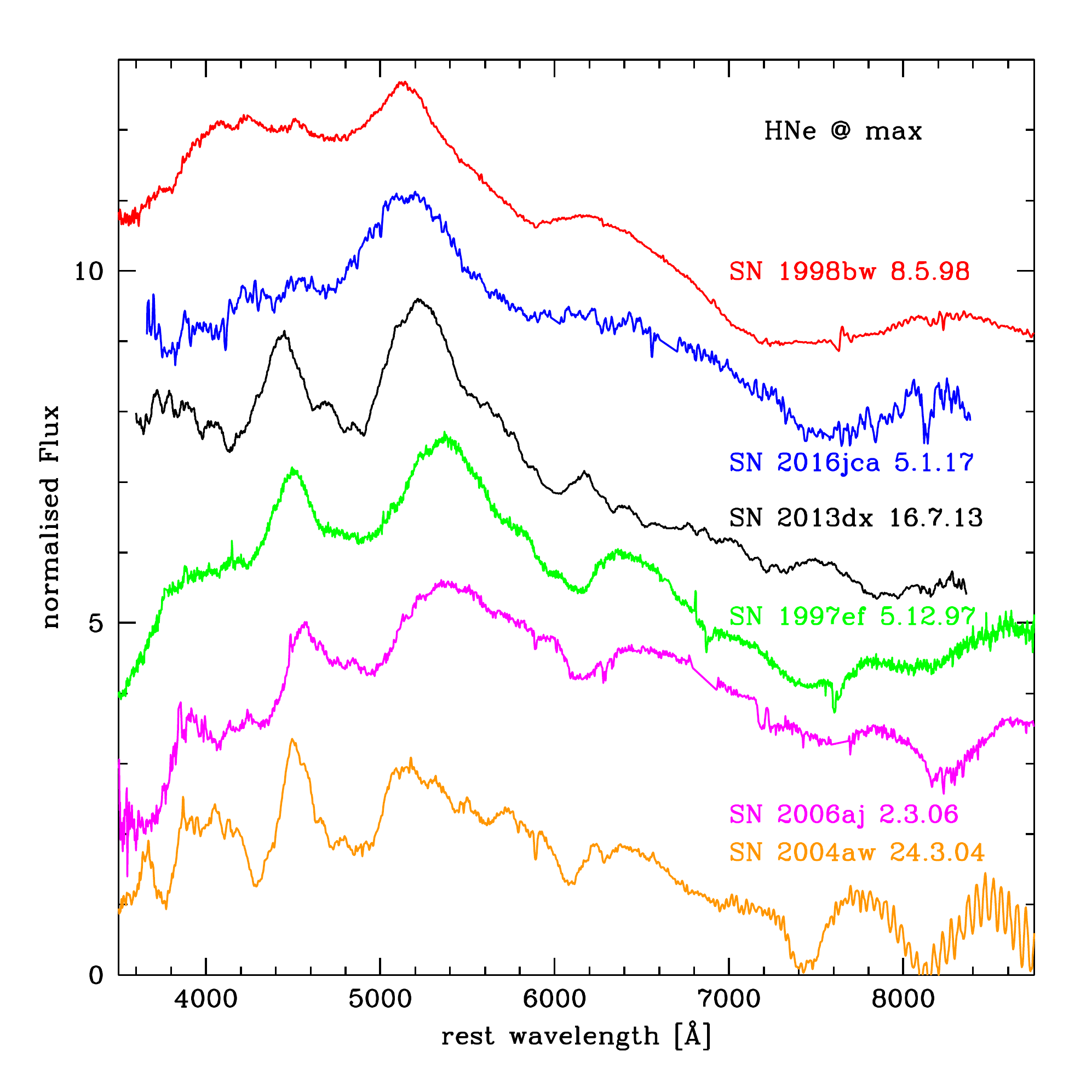}
\caption{The spectrum of SN\,2013dx taken close to maximum light compared to 
those of other broad-lined SNe\,Ic, both with and without a GRB, obtained at a 
similar epoch.}
\label{fig:speccomp_max}
\end{figure}

\begin{figure} 
\includegraphics[width=90mm]{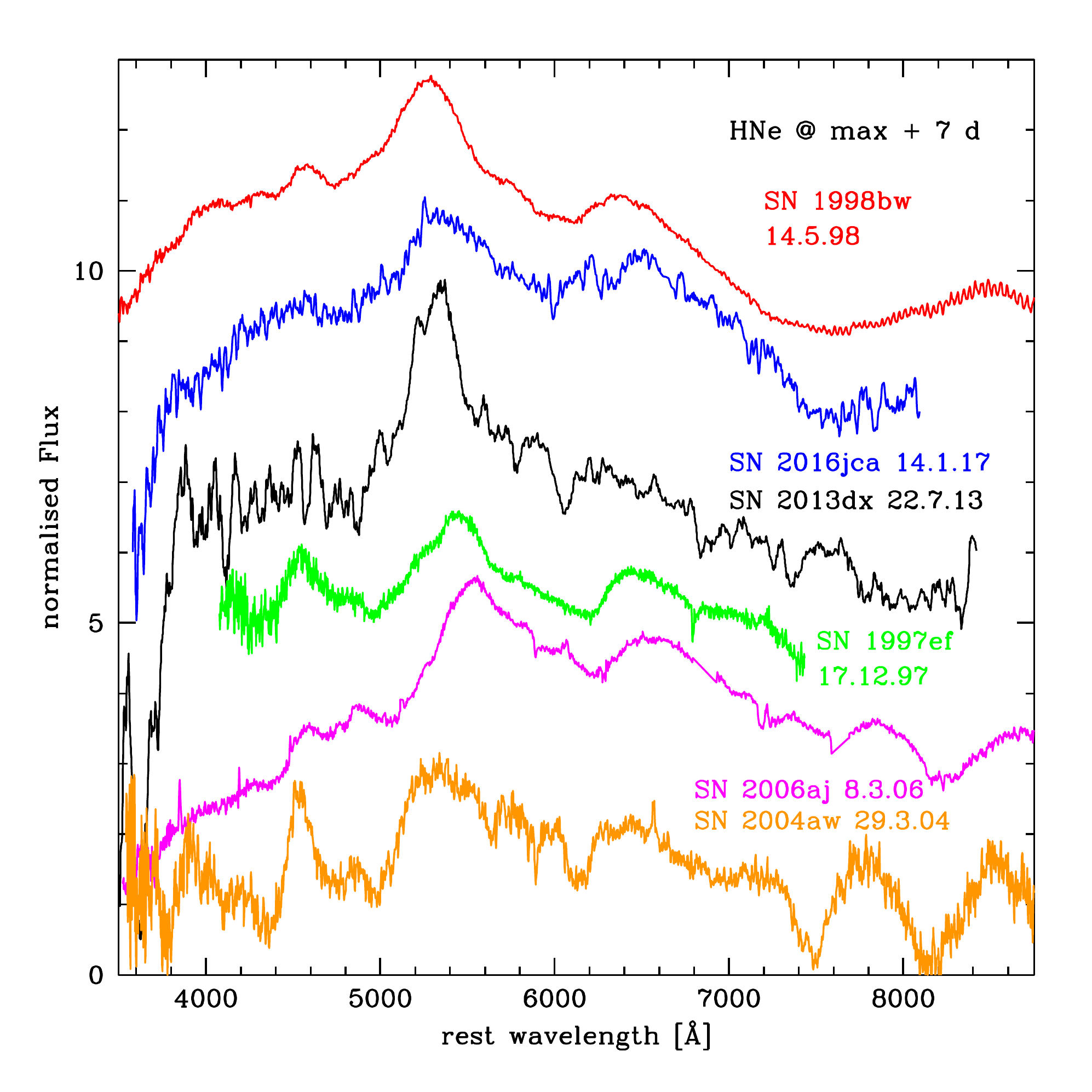}
\caption{The spectrum of SN\,2013dx taken $\approx 1$ week after maximum light 
compared to those of other broad-lined SNe\,Ic, both with and without a GRB, 
obtained at a similar epoch.}
\label{fig:speccomp_postmax}
\end{figure}

Given the complexity of establishing a density profile, we started with a
detailed comparison of the spectra of SN\,2013dx with those of other GRB/SNe as
well as broad-lined SNe\,Ic at various epochs, in order to define a closest
analogue. Figures \ref{fig:speccomp_premax}, \ref{fig:speccomp_max}, and
\ref{fig:speccomp_postmax}, show these comparisons at three epochs: $\sim 1$
week before maximum, near maximum, and $\sim 1$ week after maximum.  

At $\sim 1$ week before maximum (Fig. \ref{fig:speccomp_premax}), while the red
side of the optical spectrum of SN\,2013dx shows similarities to GRB/SNe such as
1998bw and 2016jca, the Fe-dominated absorption near 4800\,\AA\ is much deeper
and narrower, indicating a smaller velocity and a steeper density profile than
in GRB/SNe and even in SN\,1997ef. In fact, the spectrum of SN\,2013dx is not
too different from that of a lower \KE\ event such as SN\,2006aj
\citep{maz06b,pian06}. An estimate of the velocity based on the position of the
\SiII\,6347, 6371\,\AA\ line in \citet{delia15} yielded a high velocity
(26000\,\kms) at this epoch, but it must be noted that the line most probably
forms above the pseudo-photosphere, such that comparing directly-measured line
velocities and photospheric velocities obtained from modelling may be
misleading. 

\begin{figure} 
\includegraphics[width=90mm]{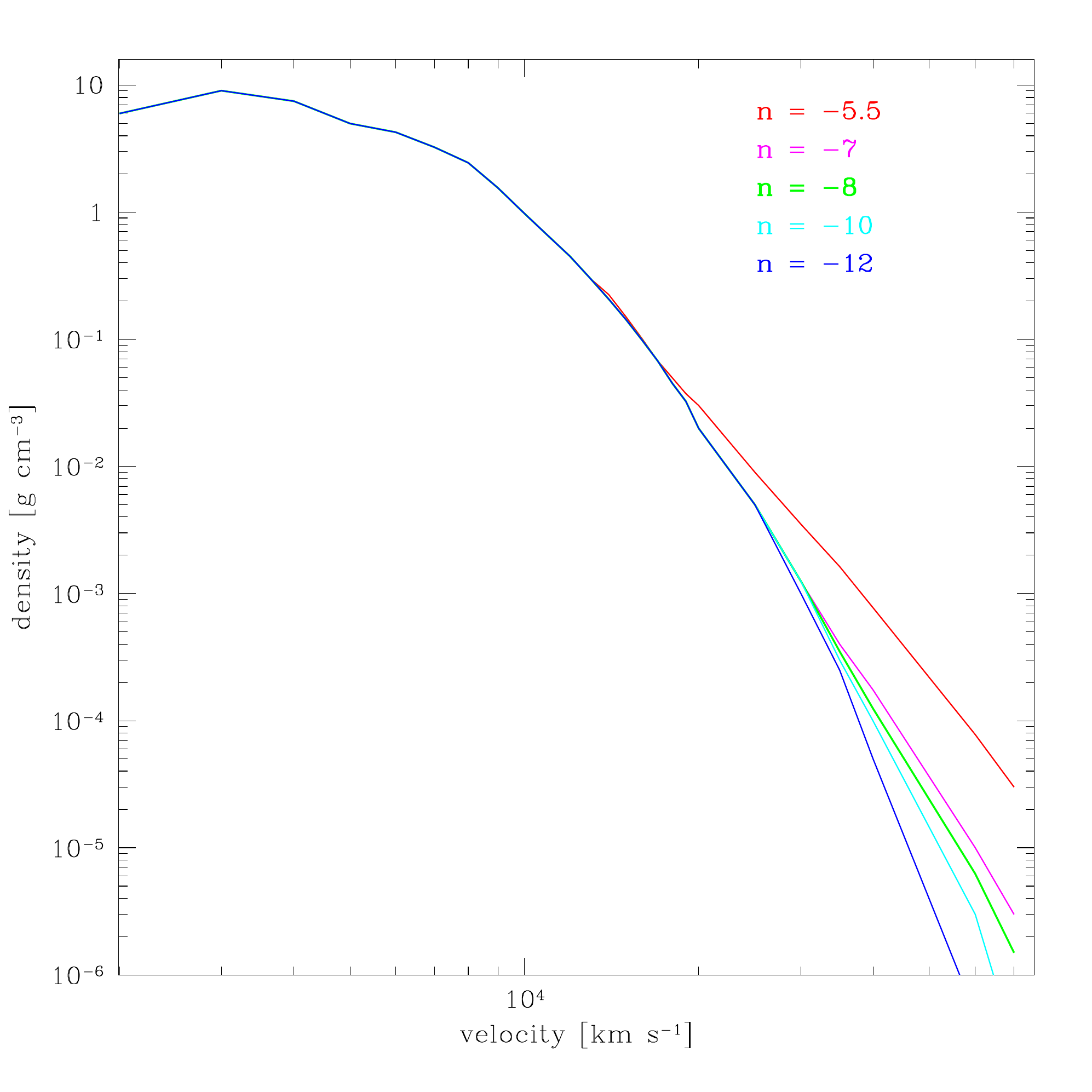}
\caption{The various density distributions that were tested to match the 
spectra of SN\,2013dx. The values of the density refer to an epoch of 100 
seconds after explosion.}
\label{fig:densities}
\end{figure}

The differences are even clearer at maximum (Fig. \ref{fig:speccomp_max}), when
a larger sample is available for comparison. SN\,2013dx looks spectroscopically
quite similar to SN\,1997ef, and not at all like GRB/SNe 1998bw or 2016jca. The
XRF/SN 2006aj is also a good match, as is the moderately energetic SN\,2004aw.
All of these SNe are characterised by significantly less line blanketing than
the GRB/SNe, and the similarity suggests that moderate line blanketing
characterises also SN\,2013dx. However, because of the relatively high redshift,
the red side of the spectrum of SN\,2013dx is not well observed. This is a
region where blending of the \CaII\,IR triplet into the \OI\,7774 line is a
clear indicator of high velocity ejecta and hence of a high \KE, so we should be
careful. A possible \SiII\,6347, 6371\,\AA\ line indicates a velocity of
$\approx 16700$\,\kms, which is significantly less than the 20000\,\kms\
reported by \citet{delia15}. The velocity may actually be even lower if the
emission near 6100\,\AA\ is spurious, as it may well be given the behaviour of
the other SNe in the comparison, which do not show much red emission in the
\SiII\ line. The observed spectra point to a significant difference in \KE\
between SN\,2013dx and classical GRB/SNe. 

Finally, the spectrum at $\approx 1$ week after maximum (Figure
\ref{fig:speccomp_postmax}) reflects the dimming of the source, and really only
shows a peak near 5200\,\AA. Additionally, the spectrum has a very low
signal-to-noise ratio, such that it is practically impossible to see a \SiII\
line, let alone use spectral features to estimate an expansion velocity. The
measurement published in \citet{delia15} is therefore likely to be highly
uncertain. 

Given these uncertainties, we tested a number of different density profiles,
characterised by a similar inner density (and therefore mass) but by different
density slopes at high velocity.  Based on the procedure followed in
\citet{maz17}, we developed density profiles with different gradients in the
outer regions. We used as a starting point a model very similar to the one used
for SN\,1998bw \citep{iwa98}, except that we reduced the mass to 9\,\Msun\ and,
accordingly, the energy to $\approx 1.4\times 10^{52}$\,erg, in line with the
expected mass of SN\,2013dx. The outer density in this model behaves like a
power law with index $n \approx -5.5$ with respect to radius.  The flat slope
was required in order to produce the broad lines seen in SN\,1998bw. As
SN\,2013dx has narrower lines, we constructed models with progressively steeper
outer density profiles, \ie\ $n \approx -7$, $-8$, $-10$, and $-12$. These
density profiles are shown in Fig. \ref{fig:densities}. While the ejected mass
does not change significantly in the various models, the value of \KE\ decreases
to $9.5\times10^{51}, 9.3\times10^{51}, 9.0\times10^{51}$, and
$8.7\times10^{51}$\,erg, respectively. 

We then computed a series of synthetic spectra for each of our density profiles.
%We modelled the VLT spectra only, as they are much better in quality than the
%other spectra presented in \citet{delia15}. 
For each epoch, we searched for the best convergence based on our choice of $L$,
\vph\ and the abundances. Because of the different radial dependencies of
density, models using different density profiles yield somewhat different
values of best-fitting \vph. The luminosity, on the other hand, changes very
little from model to model, as it mostly determines the level of the observed
flux. As for the abundances we started with the fiducial values that we used for
the outer layers of SN\,1998bw \citep{iwa98}. The initial composition in that
case was dominated by oxygen ($\approx 85$\,\% by mass), followed by neon
($\approx 10$\,\%), carbon ($\approx 2$\,\%), silicon ($\approx 1$\,\%), and
nickel ($\approx 1$\,\%). In computing the models we could fortunately fix the
value of $t$, which greatly helps defining the other parameters.  

\begin{figure*} 
\includegraphics[width=139mm]{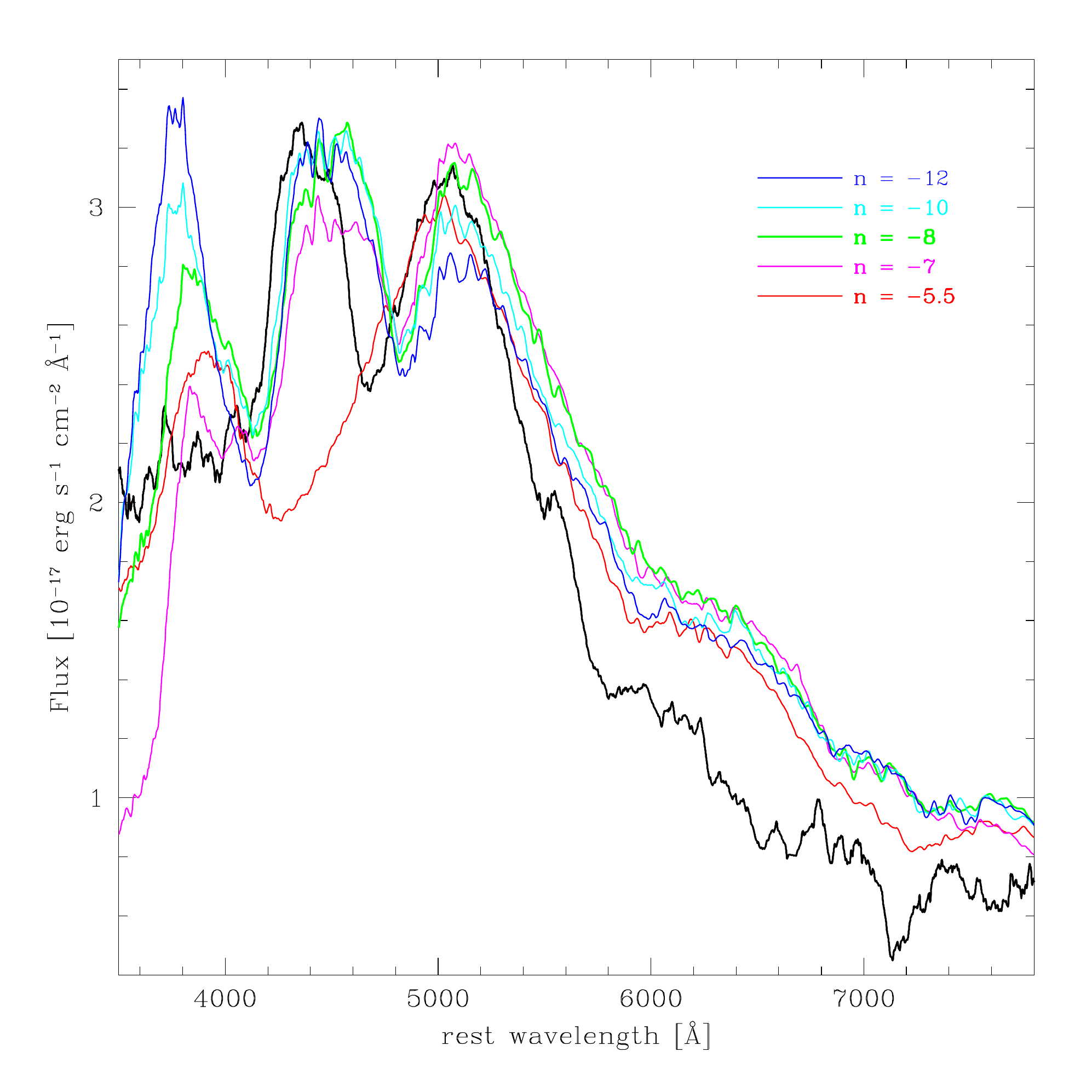}
\caption{The earliest spectrum of SN\,2013dx, obtained on 2013 July 9 at an 
epoch of 7 days, compared to synthetic spectra computed using the various 
density distributions shown in Fig. \ref{fig:densities}.}
\label{fig:densitytest}
\end{figure*}

As the earlier spectra are the most sensitive to the outer density gradient, we
show in Fig. \ref{fig:densitytest} a comparison of the synthetic spectra we
obtained for the earliest epoch available, 7 days after the GRB. We do see large
differences between the steeper and flatter models. In particular, the
low-opacity re-emission peak near 4500\,\AA, which is visible in SN\,2013dx but
absent in both SN\,1998bw and SN\,2016jca, is only reproduced in the lower
[\KE/\Mej] models. Although none of the models shown matches the observed
spectrum of SN\,2013 perfectly, better matches are provided by models with
intermediate values of the slope. Based on the results at this and other epochs,
we selected the model with $\rho \propto r^{-8}$ as our fiducial model for
SN\,2013dx.  This model has \KE$ = 9.3\times10^{51}$\,erg. 

\begin{table}
 \caption{Model parameters for the synthetic spectra. The uncertainties on the 
 input parameters can be assumed to be similar to those derived in 
 \citet{ashall20}. Following afterglow and host galaxy subtraction, the 
 uncertainty on an individual value of $L$ and $v_{\rm ph}$ is $\sim 5$\,\%. 
 However, only a specific combination of $L$ and $v_{\rm ph}$ can produce the 
 correct spectral shape and flux, demonstrating that in practice the errors are 
 much smaller than 5\,\%, and the input parameters are very well constrained.  
 Furthermore, the error on abundances can be assumed to be less than 25\% by 
 mass fraction in each shell \citep{maz08b}. }
  \label{tab:Modelinputdata}
  \centering
  \begin{tabular}{ccccc}
    Date 	& $t$  &   $L$   	 & $v_{\rm ph}$ & $T_{\rm BB}$\\
         	& [d]  & [erg/s] 	   &  [km/s]  &    [K]   \\
    \hline
     9 Jul 2013 &  7.0 & $6.03\times 10^{42}$ & 19000 & $ 8250 $ \\
    11 Jul 2013 &  8.7 & $6.61\times 10^{42}$ & 17600 & $ 8530 $ \\
    13 Jul 2013 & 10.5 & $8.13\times 10^{42}$ & 16300 & $ 8550 $ \\
    16 Jul 2013 & 13.1 & $8.32\times 10^{42}$ & 14625 & $ 8070 $ \\
    20 Jul 2013 & 16.6 & $7.41\times 10^{42}$ & 13325 & $ 7300 $ \\
    22 Jul 2013 & 18.3 & $7.00\times 10^{42}$ & 12550 & $ 7070 $ \\
    27 Jul 2013 & 22.7 & $5.69\times 10^{42}$ & 11300 & $ 6360 $ \\
    30 Jul 2013 & 25.4 & $4.90\times 10^{42}$ & 10525 & $ 6000 $ \\
     3 Aug 2013 & 28.8 & $4.03\times 10^{42}$ &  9350 & $ 5700 $ \\
     6 Aug 2013 & 31.5 & $3.27\times 10^{42}$ &  8475 & $ 5440 $ \\
    10 Aug 2013 & 35.0 & $2.09\times 10^{42}$ &  7100 & $ 5240 $ \\
    \hline
  \end{tabular}
\end{table}

The synthetic spectrum obtained with that model matches the observed one reasonably well, in particular with regard to reproducing the re-emission peak near 4400\,\AA. As Fig. \ref{fig:densitytest} shows, none of the models reproduces the blueshift of the \FeII\ feature near 4700\,\AA, and it is difficult to match the flux near 4000\,\AA, where however the data may be uncertain. The density profile we selected offers the best match overall.

\begin{figure*} 
\includegraphics[width=139mm]{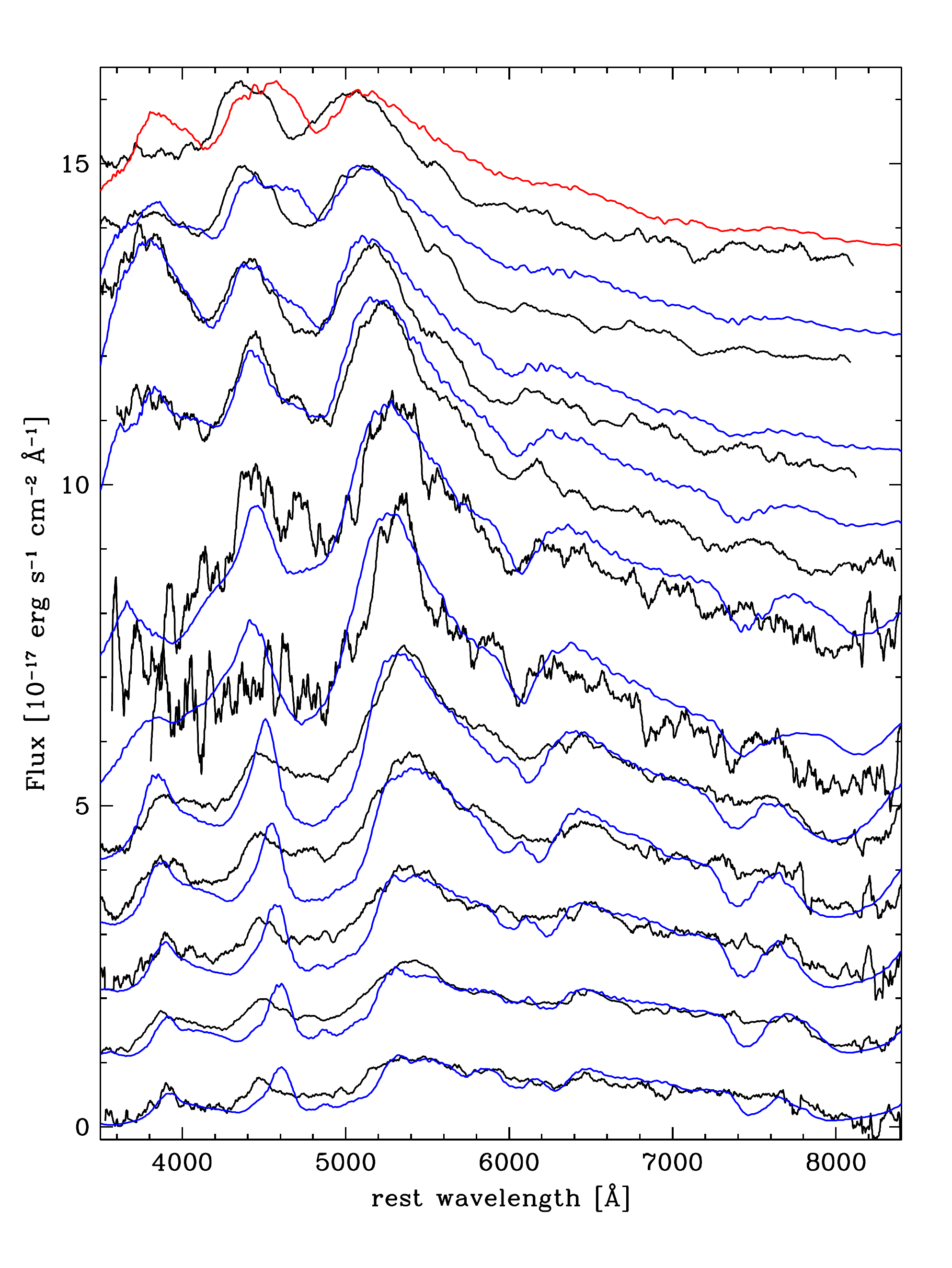}
\caption{Synthetic spectral fits for the time series of VLT spectra obtained using the density profile with outer density $\rho \propto r^{-8}$. The spectra are displayed from top to bottom in the order shown in Table \ref{tab:Modelinputdata}. The flux has been increased by constant values for displaying purposes.}
\label{fig:specfits}
\end{figure*}

Having now selected a favoured model, we proceeded with modelling the time series of VLT spectra. We performed modelling for all density profiles, but here we show only the results for the profile we selected (Fig. \ref{fig:specfits}). In computing the models we evolved the density to the appropriate time, used the best-fitting values of $L$ and \vph\ (see values in Table \ref{tab:Modelinputdata}), and adapted the abundances to optimise the fit. The abundances did not change much as a function of depth, although we did see a slightly decreasing abundance of \Nifs\ with decreasing radius, which confirms that a higher fraction of \Nifs\ can be located at high velocities near the direction of the jet, because of the combination of abundance and area sampled \citep[see][]{ashall19}. Overall, the dominant element in the outer layers of the ejecta that are sampled by the spectra is oxygen ($\approx 75$\,\% by mass), followed by neon ($\approx 14$\,\%), nickel ($\approx 5$\,\%), iron ($\approx 2$\,\%), and silicon ($\approx 1$\,\%). The synthetic spectra reproduce the observed spectral evolution reasonably well, including line broadening. As time progresses and the degree of ionization decreases, the mismatch in the \FeII\ trough near 4700\,\AA\ disappears. This is a consequence of the lower ionization, which means that the \FeII\ lines become stronger, while at the earliest epochs \FeIII\ is more dominant and the \FeII\ lines only form at very high velocities, such that they actually contribute to the next absorption feature, near 4200\,\AA. The \SiII\,6355 line becomes stronger with time as the degree of ionization decreases. The \CaII\,IR triplet is isolated from the \OI\,7774 line, but the feature is weak and falls partly outside of the spectral region available because of the redshift of SN\,2013dx, so that the calcium abundance is not easy to determine.

The model we selected has \Mej\,$= 9$\,\Msun\, and \KE\,$= 9 \times 10^{51}$\,erg.  Of course our results are approximate. The neighbouring models that we tested give a sense of the effect \KE\ has on the spectra. We therefore conservatively estimate that \Mej\,$= 9\pm2$\,\Msun\, and \KE\,$= 9\pm2 \times 10^{51}$\,erg.

\section{Light curve model} 
\label{sec:LC}

Having selected a plausible explosion model, one way to verify that it is viable
is to use it to compute a synthetic light curve and test it against the observed
one. 

\begin{figure*} 
\includegraphics[width=139mm]{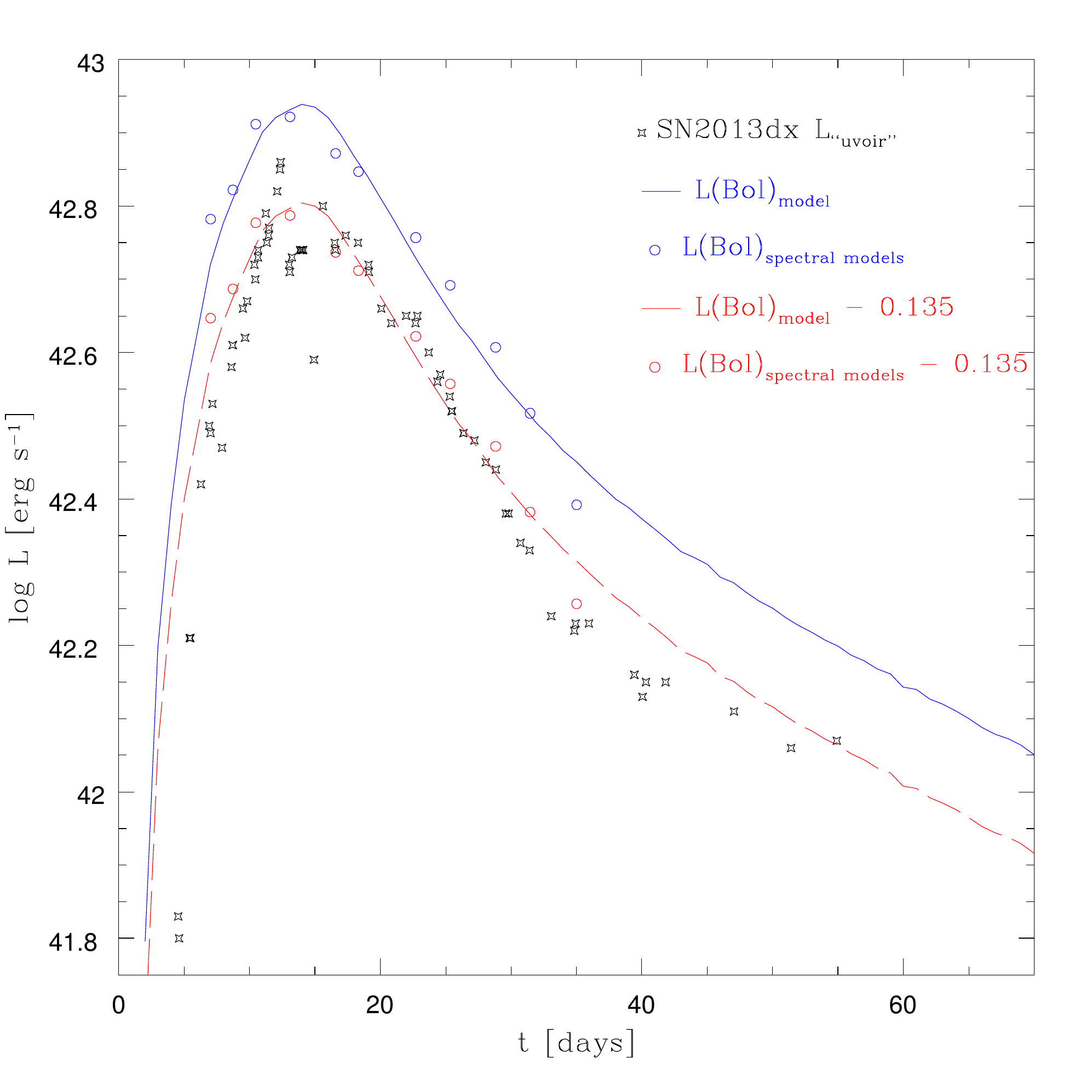}
\caption{The synthetic bolometric light curve obtained using the density 
profile and abundances derived from our modelling (blue line), compared to the 
pseudo-bolometric light curve of SN\,2013dx computed in the interval 
4500-10500\,\AA\  (black stars). The luminosities that were used in the 
spectral models are shown as blue circles. The red dashed line and the red 
circles are the synthetic bolometric light curve and the luminosities used in 
the spectral models shifted downwards by 0.135\,dex.}
\label{fig:LCfit}
\end{figure*}

Using the density and abundance structures that we determined from spectral
modelling, we used our SN Montecarlo light curve code to compute a synthetic
bolometric light curve. The code was initially presented in \citet{capp97}, and
then further developed and used in \citet{maz01a} and later papers. Using a
radial distribution of abundances and densities in homologously expanding
ejecta, the code uses a Montecarlo method to compute the deposition of gamma
rays and positrons emitted in the radioactive decay chain of \Nifs. Both
gamma-rays and positrons are allowed to propagate within the ejecta, where they
deposit their energy. Their propagation is subject to opacities $\kappa_{\gamma}
= 0.027$\,g\,cm$^{-2}$ for gamma-rays and $\kappa_{e^+} = 7$\,g\,cm$^{-2}$ for
positrons, respectively. Both gamma-rays and positrons are assumed to deposit
their energy in its entirety when they are deemed to have encountered a
sufficiently large optical depth in their propagation.  The energy thus
deposited is converted into optical photon energy. Energy packets that represent
optical photons are themselves allowed to propagate within the ejecta, and are
subject to a scattering opacity whose value is determined based on the
abundances within the ejecta \citep{maz01a}. This approach is based on the
assumption that in hydrogen-free ejecta line opacity is the dominant form of
opacity. In order to conserve energy, every absorption process is followed by an
emission one, so that photons only change direction, and thus experience
different travel lengths and times before they finally escape. This transforms
the exponential energy input from radioactivity into the typical rising and
declining shape of a Type I SN light curve. The code has been used to model the
light curves of many Type I SNe, both Ia and Ib/c \citep[\eg][]{maz17,teffs20}.

Given that we could not use nebular spectra to determine the properties of the
inner ejecta, including densities and abundances, the light curve modelling is
not completely self-consistent. The mass of \Nifs\ was to some extent a free
parameter, which we adjusted in order to optimise the fit. The region of the
ejecta that we sampled with the spectral modelling extends down to a velocity of
$\approx 7000$\,km/s. This includes a mass of $\approx 6$\,\Msun, leaving some
3\,\Msun\ of the inner ejecta unexplored. Within the sampled mass we already
include some 0.32\,\Msun\ of \Nifs, with an abundance distribution that
decreases slowly with velocity. In the inner, unsampled part of the ejecta, we
fixed the \Nifs\ fraction to the same low level as in the innermost layers that
were sampled, $\sim 2$\%. This yields a total \Nifs\ mass of 0.38\,\Msun, and
produces the light curve shown in Fig. \ref{fig:LCfit}. 

When compared to the computed pseudo-bolometric light curve, the synthetic one
lies about 0.135\,dex above, which is similar to the difference between the
pseudo-bolometric light curve of SN\,1998bw computed over the same wavelength
range as that of SN\,2013dx (4500-10500\,\AA), and that computed over the
broader range 3650-22200\,\AA.  This discrepancy is almost certainly due to
unobserved near-ultraviolet (NUV) and, especially, near-infrared (NIR)
contributions, as discussed in Sect. \ref{sncomponent}.   In Fig.
\ref{fig:LCfit} we also show the synthetic light curve scaled down by
0.135\,dex: the good match between the model and the observations seems to
suggest that our estimate of the abundance of \Nifs\ in the innermost layers is
roughly correct. Even if we assumed that no \Nifs\ is present below 7000\,\kms,
the synthetic light curve would still lie above the observed one. 

\begin{figure*} 
\includegraphics[width=139mm]{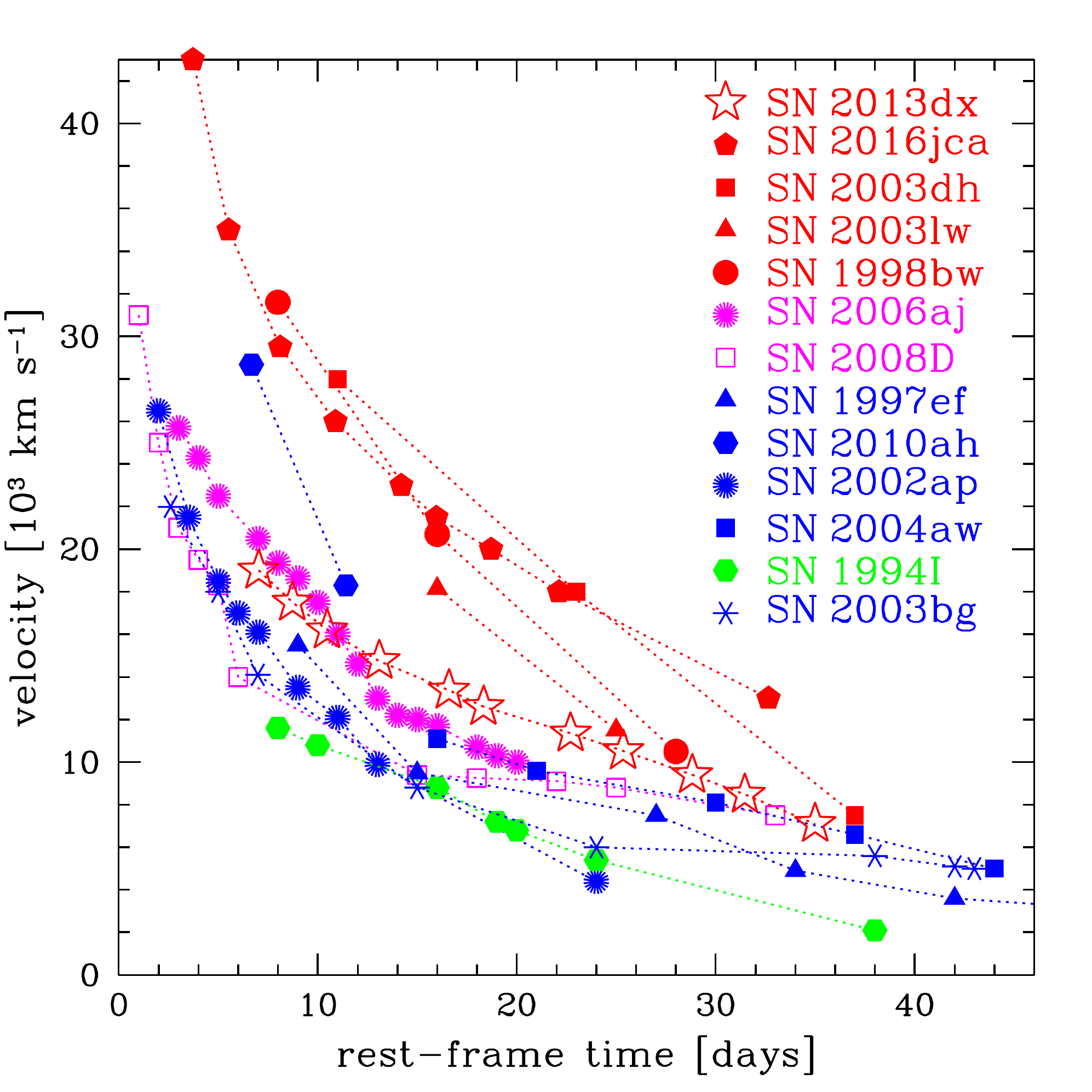}
\caption{The evolution of the velocity of the pseudo-photosphere as determined 
by our model spectra for SN\,2013dx and a number of broad-lined 
stripped-envelope SNe.}
\label{fig:velocities}
\end{figure*}

\section{Discussion}
\label{sec:Disc}

The properties of SN\,2013dx make it quite an unusual GRB/SN. Although it does
have broad lines, it is not nearly as broad-lined as the other GRB/SNe. The
breadth of the lines is normally associated with the ratio \KE/\Mej. In the
context of the scheme proposed by \citet{prenticemazz18}, SN\,2013dx would be
classified as a SN\,Ic-4 based upon the number of spectral features in the
optical range. GRB/SNe are all Ic-3. The smaller line blending is the result of
a lower \KE. We derive for SN\,2013dx a value \KE$\approx 10^{52}$\,erg. In
GRB/SNe, the value of \KE\ derived from one-dimensional modelling is thought to
be overestimated by factors of 2-3 because of the intrinsic asphericity of the
explosion \citep{maz05,mae02}. As GRB/SNe are viewed on or near the axis of the
GRB jet, we also tend to pick up the highest velocity SN ejecta, which may lead
to an overestimate of \KE\ if a one-dimensional analysis is performed.
The actual value of \KE\ for SN\,2013dx may therefore be of the order of
$3-5\,\times\,10^{51}$\,erg, depending on the actual degree of asphericity of
the ejecta. This energy would make SN\,2013dx not dissimilar to SNe such as
2004aw or 2010ah. Unfortunately, lack of nebular lines detection makes an
estimate of the ejecta asphericity very difficult. Yet, even if the asphericity
of SN\,2013dx was significantly less than what was inferred in - say -
SN\,1998bw \citep{maz01b}, its kinetic energy would be unlikely to be much more
than $10^{52}$\,erg. 

Based on this information, we can try to place SN\,2013dx in the context of
other SNe\,Ic, with and without a GRB. First of all, a carbon-oxygen core mass
of $\sim 9$\,\Msun\ is similar to that of other GRB/SNe. Therefore, the
progenitor star must have also been similarly massive, probably in the range
35-50\,\Msun. In this sense SN\,2013dx does not constitute an exception, which
is reassuring, as a large mass remains an essential factor in causing GRB/SNe.

First, we plot the evolution of the photospheric velocity as derived from our
models (Fig. \ref{fig:velocities}). This shows that the \vph\ of SN\,2013dx was
always significantly smaller than that of all other GRB/SNe, and indeed much
closer to several non-GRB broad-lined SNe\,Ic. 

\begin{figure} 
\includegraphics[width=89mm]{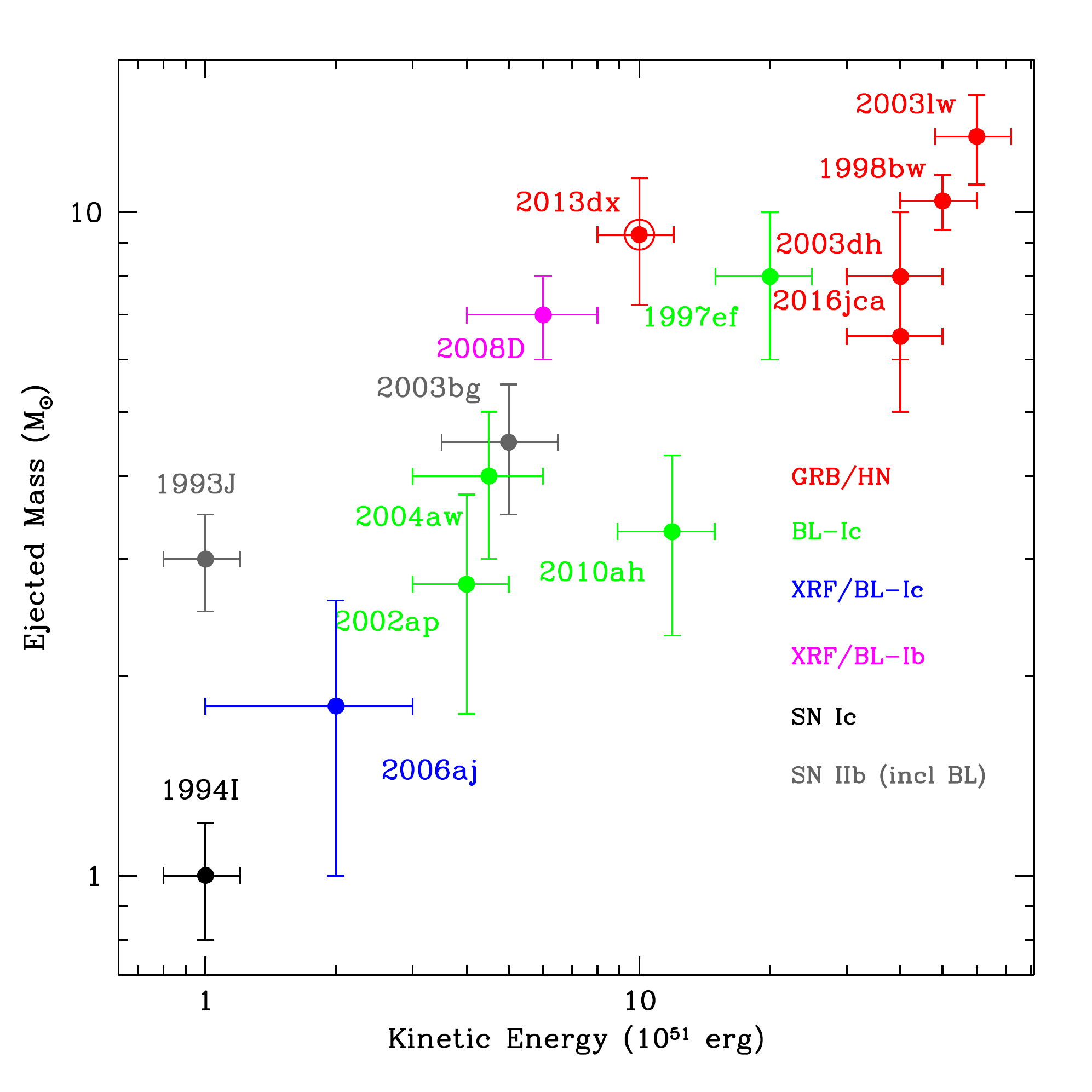}
\caption{Ejected mass vs. kinetic energy in SN\,2013dx (highlighted) and in 
other stripped-envelope SNe.  As shown in the figure, red dots indicate 
GRB/SNe, green dots non-GRB broad-lined SNe\,Ic, black dots narrow-lined 
SNe\,Ic (SN\,1994I), blue dots XRF/SNe\,Ic (SN\,2006aj), purple dots 
XRF/SNe\,Ib (SN\,2008D), grey dots SNe\,IIb. }
\label{fig:EkMej}
\end{figure}

\begin{figure} 
\includegraphics[width=89mm]{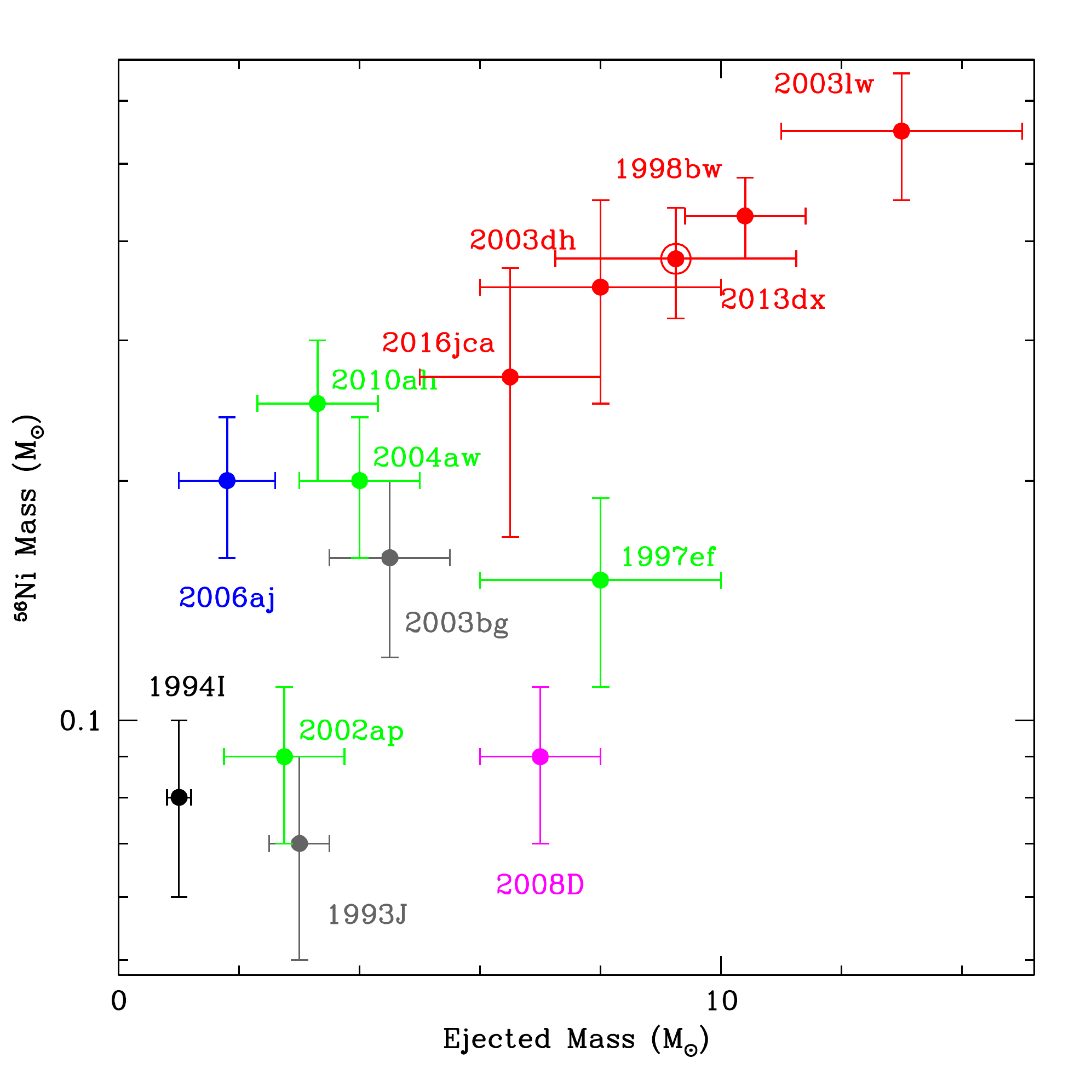}
\caption{\Nifs\ mass vs. ejected mass in SN\,2013dx (highlighted) and in other 
stripped-envelope SNe. Colour coding as in Fig. \ref{fig:EkMej}.}
\label{fig:NiMej}
\end{figure}

\begin{figure} 
\includegraphics[width=89mm]{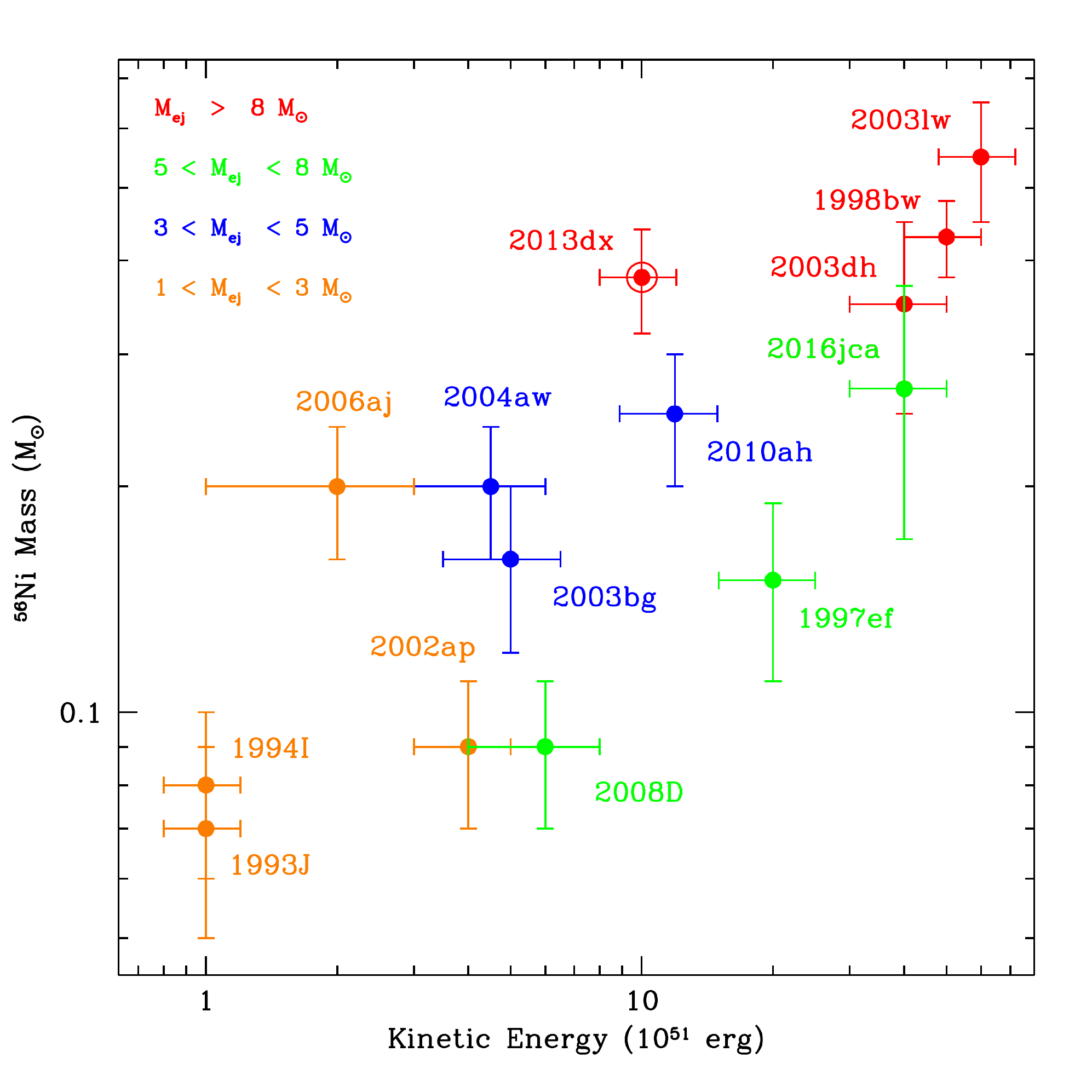}
\caption{\Nifs\ mass vs. kinetic energy in SN\,2013dx (highlighted) and in 
other stripped-envelope SNe. The colour coding is based on the ejected mass and 
is explained in the figure legend.}
\label{fig:EkNi}
\end{figure}

\begin{figure} 
\includegraphics[width=89mm]{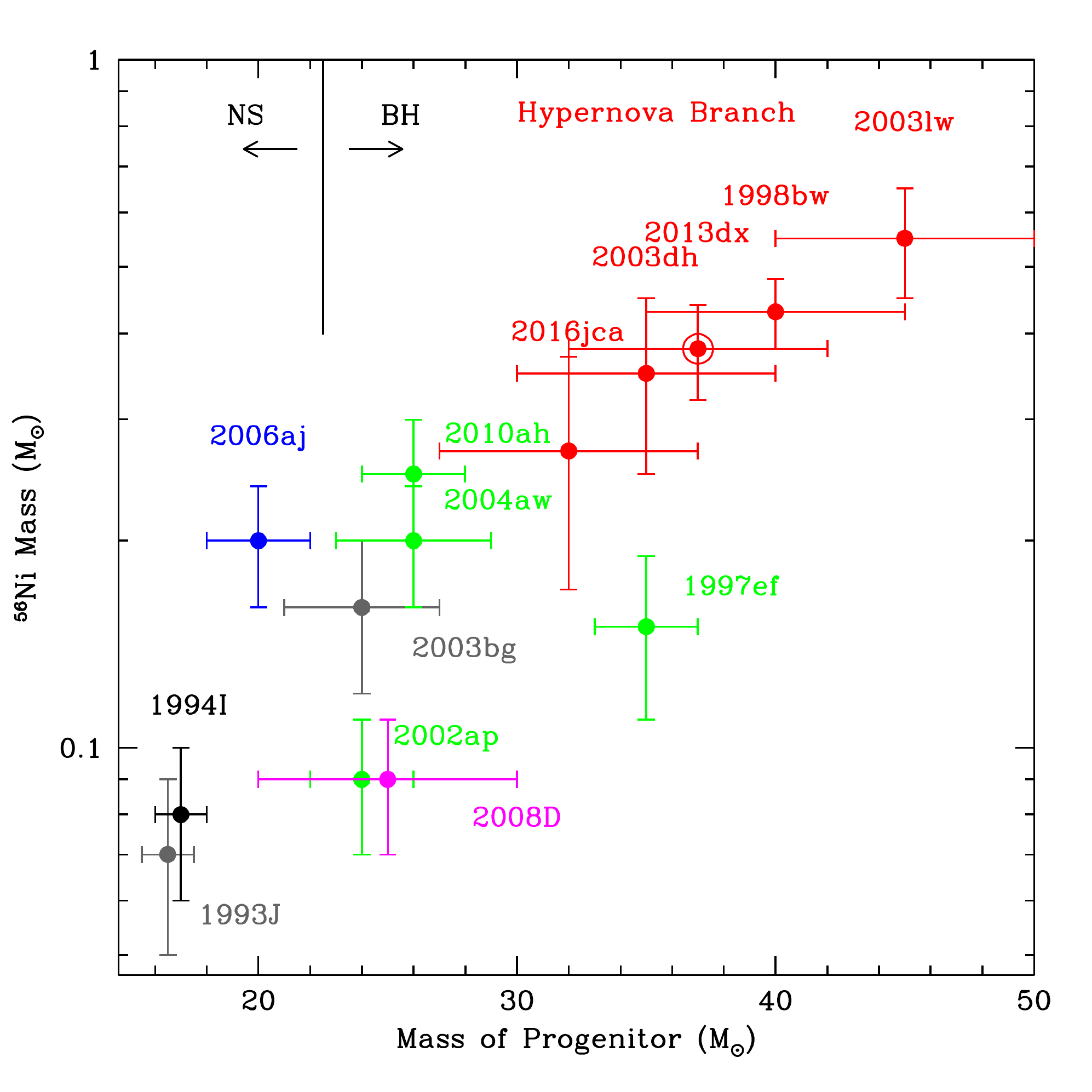}
\caption{\Nifs\ mass vs. inferred progenitor mass in SN\,2013dx (highlighted) 
and stripped-envelope in other SNe\,Ic. Colour coding as in Fig. 
\ref{fig:EkMej}.}
\label{fig:NiMprog}
\end{figure}

\begin{figure} 
\includegraphics[width=89mm]{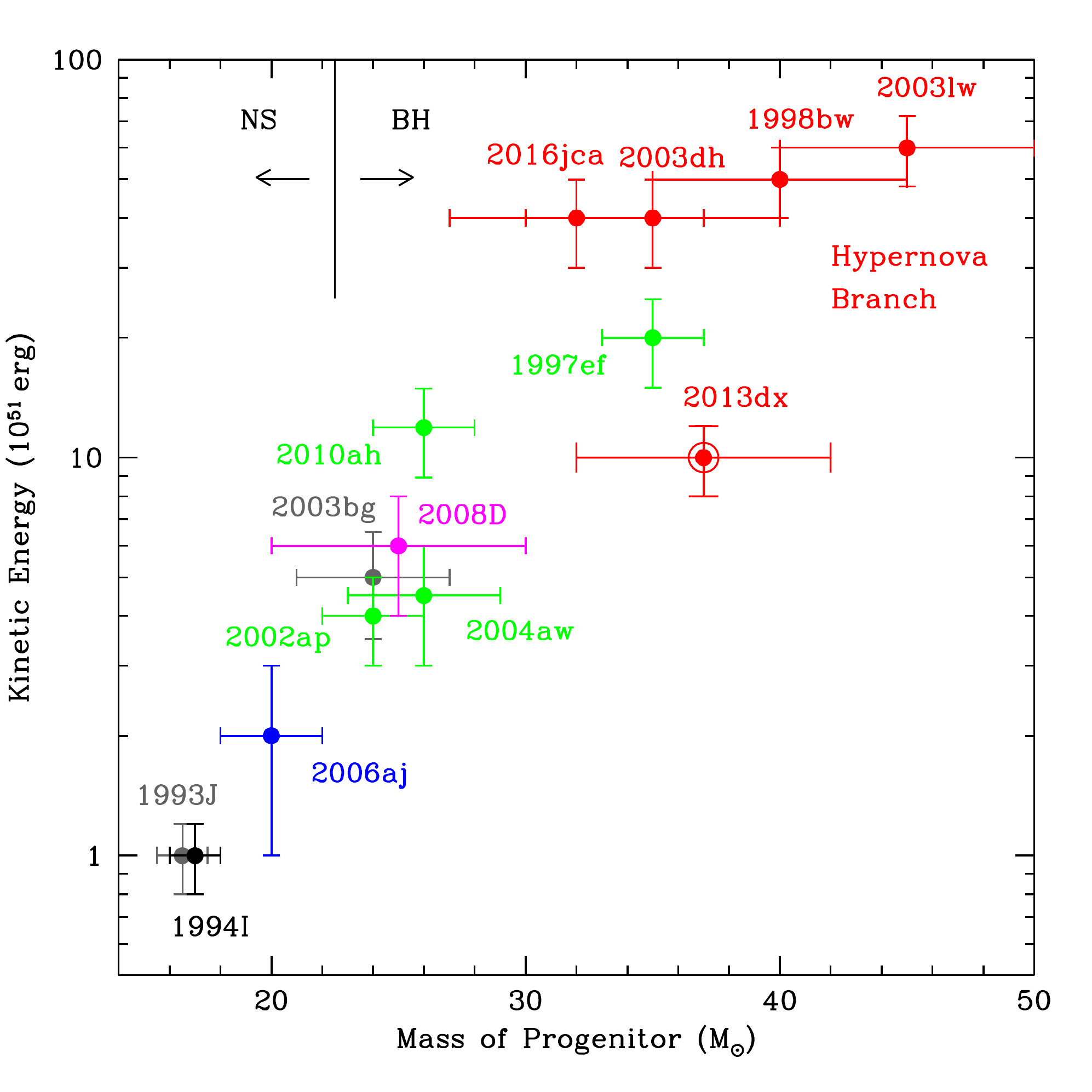}
\caption{Kinetic energy vs. inferred progenitor mass in SN\,2013dx 
(highlighted) and stripped-envelope in other SNe\,Ic. Colour coding as in 
Fig. \ref{fig:EkMej}}
\label{fig:EkMprog}
\end{figure}

We then show some plots where the main properties of a number of
stripped-envelope SNe (\Mej, \KE, M(\Nifs), M(prog)) are compared to one another
(Figures \ref{fig:EkMej}, \ref{fig:NiMej}, \ref{fig:EkNi}, \ref{fig:NiMprog},
\ref{fig:EkMprog}).   We only plot SNe for which the results were obtained
with our method, as a simpler approach based on light curve shape and a single
velocity measurement is likely to yield inconsistent results. The position of
SN\,2013dx is highlighted. While SN\,2013dx falls in the group of GRB/SNe in
plots that involve \Mej, M(prog), and M(\Nifs), it is separated from them in
terms of \KE. What is particularly interesting is the plot of \KE\ versus
M(\Nifs) (Fig. \ref{fig:EkNi}).  SN\,2013dx synthesised a similar amount of
\Nifs\ as other GRB/SNe, but had a much smaller \KE. This seems to indicate that
the relation between \KE\ and M(\Nifs) is not direct. At the same time, if we
look at the plots involving M(\Nifs) and either the ejected mass or the inferred
progenitor mass (Figs. \ref{fig:NiMej} and \ref{fig:NiMprog}), we see that
SN\,2013dx falls nicely along the trend that is defined by the other
well-studied cases, suggesting that there is a strong correlation between
progenitor mass and mass of \Nifs\ synthesised in the explosion.

The main results we obtained for SN\,2013dx are: M(\Nifs)$\approx 0.38$\,\Msun;
\Mej$\approx 9 \pm 2$\,\Msun; \KE$\approx 9 \pm 2 \times 10^{51}$\,erg. Apart
from the ejected mass, which is the parameter that can be obtained most reliably
using the width of the light curve, our results differ significantly from those
of \citet{delia15}. The difference in the mass of ejected \Nifs\ can be mostly
attributed to our new treatment of the photometry and the bolometric light curve
of SN\,2013dx. The difference in kinetic energy, for which we find a much lower
value, must however be ascribed to the fact that \citet{delia15} used a single
measurement of the velocity near light curve peak, thus missing the fact that at
early times the narrower-lined character of the spectra indicates a lack of
high-velocity ejecta, which actually carry most of the energy in a classical
GRB/SN like SN\,1998bw.

\section{Conclusions}
\label{sec:Conclusion}

\citet{maz14} suggested that the explosion kinetic energy of GRB/SNe  
\citep[not the light curve, as suggested by][]{woosley10,kasbild10} is driven by
the energy of a magnetar. The similar luminosity of all GRB/SNe may suggest that
M(\Nifs) is also related to the energy in a magnetar. It may be that some of the
rotational or magnetic energy of the magnetar contributed to the explosion, and
some to the nucleosynthesis. Nucleosynthesis requires very high densities, so it
is possible that when magnetar energy is first released, presumably soon after
the end of the collapse phase, just after the proto-neutron star is formed and
before it collapses to a black hole, nucleosynthesis is the first result. Later,
as the outer layers of the collapsing star begin to expand and densities drop,
magnetar energy can only be used to energise the explosion. While all
well-studied GRB/SNe so far reached comparable \KE\ and produced comparable
amounts of \Nifs, SN\,2013dx had a smaller \KE. Some possibilities are that
initially (in the first fraction of a second, which we may call the ``burning''
phase), the proto-NS heated the ejecta causing nucleosynthesis, but that in the
later phase, for roughly another second, the ``acceleration'' phase, the
magnetar ran out of energy to expend on accelerating the ejecta. This may be
because it had less energy overall, or it released it more rapidly, or simply
that at that point it collapsed to a BH and was no longer able to release
energy. It is also interesting that the inner layers are not expected to contain
much \Nifs. This may also be a result of different conditions at the time of
collapse than in other GRB/SNe. 

Clearly these are just suggestions, but even if they were true the fact that the
GRB was rather normal should also not be very surprising. From a jet break at 3
days, we obtain, for a CSM density of 1 cm$^{-3}$ and an observed isotropic
energy of $6 \times 10^{50}$ erg  \citep[rescaled from the original value
reported in][ using $H_0 = 73$\,km\,s$^{-1}$\,Mpc$^{-1}$)]{amati2013}, a jet
total opening angle of 20 deg \citep{sari1999}, which implies a jet corrected
energy of $\sim 10^{49}$ erg.  The jet that causes the GRB would be produced
when the proto-NS collapses to a BH, and the conditions for that process would
have been no different in SN\,2013dx than in other GRB/SNe, so that the energy
carried by the GRB is a rather small fraction of the SN kinetic energy also in
this case. Certainly this field is holding many secrets yet, and only continued
investigation, both observational and theoretical, can finally reveal them all.

\section*{Acknowledgments} 
The late time observations were performed under ESO programme 092.D-0043(A).

\section*{Data availability}
The photometric and spectroscopic data presented in this article are
publicly available via the Weizmann Interactive Supernova Data
Repository,  at https://wiserep.weizmann.ac.il.

\end{document}